\documentclass[10pt, conference, compsocconf]{IEEEtran}
\newcommand{\ignore}[1]{}
\usepackage{graphicx}
\usepackage{fancyhdr}
\usepackage[normalem]{ulem}
\PassOptionsToPackage{hyphens}{url}

\usepackage[bookmarks=false]{hyperref}

\usepackage[normalem]{ulem}
\usepackage{siunitx}

\usepackage{booktabs}
\usepackage{multirow}

\usepackage{subfig}

\ifCLASSINFOpdf
\else
\fi
\begin{document}
%
\title{Near-Memory Address Translation}



\author{
	Javier Picorel \textsuperscript{\normalsize \dag} 
	\hspace{0.15in}
	Djordje Jevdjic \textsuperscript{\normalsize \ddag} \hspace{0.15in}
	Babak Falsafi \textsuperscript{\normalsize \dag}\\ \\
	\textsuperscript{\normalsize \dag}\textit{EcoCloud, EPFL} \hspace{.3in}
	\textsuperscript{\normalsize \ddag}\textit{Microsoft Research} \hspace{.3in}
	\vspace{-0.01in}
}


%


\maketitle

\sloppy

\begin{abstract}

Memory and logic integration on the same chip is becoming increasingly cost effective, creating the opportunity to offload data-intensive functionality to processing units placed inside memory chips. The introduction of memory-side processing units (MPUs) into conventional systems faces virtual memory as the first big showstopper: without efficient hardware support for address translation MPUs have highly limited applicability. Unfortunately, conventional translation mechanisms fall short of providing fast translations as contemporary memories exceed the reach of TLBs, making expensive page walks common.



In this paper, we are the first to show that the historically important flexibility to map any virtual page to any page frame is unnecessary in today's servers. We find that while limiting the associativity of the virtual-to-physical mapping incurs no penalty, it can break the translate-then-fetch serialization if combined with careful data placement in the MPU's memory, allowing for translation and data fetch to proceed independently and in parallel. We propose the Distributed Inverted Page Table (DIPTA), a near-memory structure in which the smallest memory partition keeps the translation information for its data share, ensuring that the translation completes together with the data fetch. DIPTA completely eliminates the performance overhead of translation, achieving speedups of up to $3.81\times$ and $2.13\times$ over conventional translation using 4KB and 1GB pages respectively.

\end{abstract}


\section{Introduction}


\begin{figure*}
\centering
\includegraphics[width=0.90\textwidth]{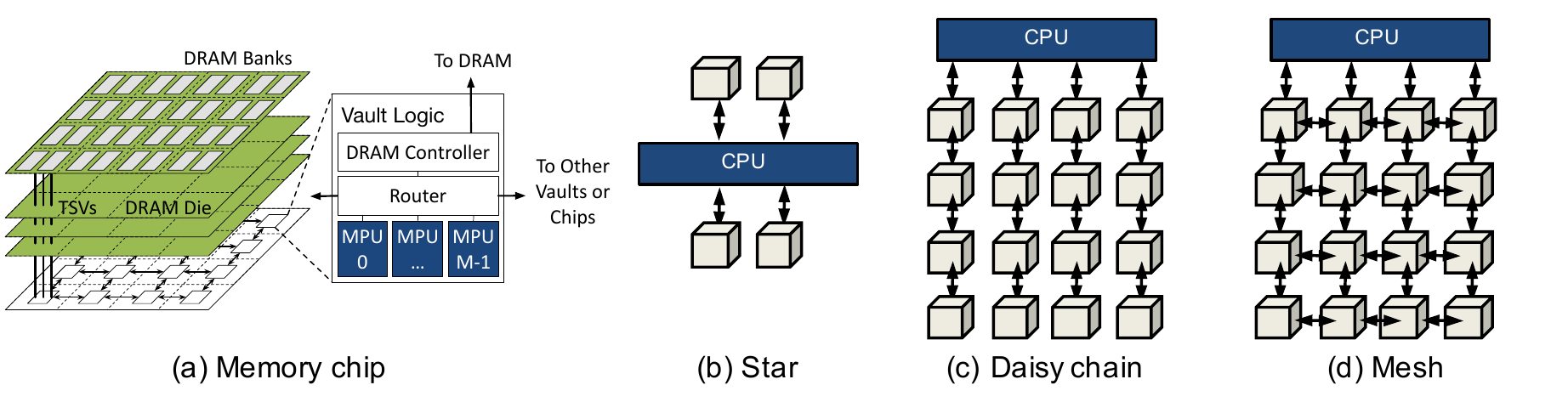}
\caption{Overview of an MPU chip and memory network topologies proposed in the literature.}
\label{fig:overview}
\end{figure*}

Recent advancements in die-stacking technology have enabled the integration of logic into conventional DRAM chips~\cite{hmc, diram}, reviving the decades old idea of processing in memory~\cite{draper:architecture, kang:flexram, kozyrakis:scalable, oskin:active}. The logic in memory devices can leverage the physical data proximity and immense internal bandwidth to perform memory-intensive functionalities. This computation paradigm is known as \emph{near-memory processing}, and we refer to the computing devices as \emph{near-Memory Processing Units} (MPUs). 




MPU's adoption into conventional systems faces virtual memory (VM) as the first major setback; without an address translation mechanism, the usability of MPUs is severely limited~\cite{power:supporting, vesely:observation}. Unfortunately, equipping MPUs with conventional translation hardware comes at a high performance overhead. Limited TLB reach causes high TLB miss rates that grow with the dataset size and memory capacity~\cite{basu:efficient,karakostas:redundant}. The TLB miss penalty also grows with the memory size as page tables can reside in any memory chip. 


In this work, we leverage the observation that most modern online services and analytic engines are practically memory resident~\cite{oracle:timesten, sap:hana, basu:efficient, bronson:tao, ferdman:clearing, haque:few-to-many, kozyrakis:server, ousterhout:case, power:implications, zaharia:spark}. These systems experience page swapping extremely infrequently and have much less fragmented memory layouts. Because contiguous virtual pages are often mapped to contiguous physical pages~\cite{pham:increasing, pham:colt}, the conventional fully associative page placement flexibility is overkill. We are the first to show that restricting the mapping between virtual to physical addresses from fully associative to set associative (or direct mapped) has practically no impact on the page fault rate. 




Based on this novel observation, we propose the Distributed Inverted Page Table (DIPTA) translation mechanism. DIPTA restricts the associativity so that all but a few bits remain invariant across the virtual-to-physical mapping, and with a highly accurate way predictor, the unknown bits are figured out so that address translation and data fetch are completely independent. Furthermore, to ensure that the data fetch and translation are completely overlapped, we place the page table entries next to the data in the form of an inverted page table, either in SRAM or embedded in DRAM. Overall, DIPTA completely eliminates the ever-growing performance overhead of translation for in-memory workloads.

The primary contributions of this paper are:

%
%
%
%

\noindent $\bullet$  We show that address translation in the case of MPUs suffers from limited reach and increasingly high miss penalty, and can dramatically increase the execution time by more than $3\times$.

\noindent $\bullet$ We show that modern server workloads do not need a fully associative VM and can tolerate associativity ranging from direct-mapped to 4-way associative.

\noindent $\bullet$ We propose DIPTA, a scalable near-memory address translation mechanism. DIPTA leverages the limited associativity, which combined with a novel data placement and highly-accurate near-memory way prediction, enables MPUs to fully overlap address translation with data fetch.

\noindent $\bullet$ We propose two DIPTA implementations; one simple in SRAM and one in DRAM. The DRAM implementation performs data fetch and address translation in a single DRAM access, and presents a novel DRAM layout to embed metadata without affecting the OS page size.

Using a combination of trace-driven functional and full-system cycle-accurate simulation, we show that DIPTA eliminates the address translation overhead, providing speedups of up to $3.81\times$ and $2.13\times$ over conventional translation using 4KB and 1GB pages respectively. Our proposed $256$B way predictor  (per memory partition) achieves $69$\%-$91$\% coverage for worst-case workloads.

The rest of this paper is organized as follows. Section~\ref{sec:bg} introduces background on near-memory architectures and virtual memory. Section~\ref{sec:vm} presents the associativity requirements of in-memory server workloads. Sections~\ref{sec:design} and \ref{sec:system} describe the DIPTA design and discuss the system-level implications, respectively. Section~\ref{sec:evaluation} presents the evaluation methodology and results. Section~\ref{sec:related} discusses the related work, and Section~\ref{sec:conclusion} concludes the paper.





\section{Background}
\label{sec:bg}
\subsection{Near-Memory Architectures}


Figure~\ref{fig:overview} (a) illustrates the anatomy of an MPU chip. We assume an organization similar to JEDEC's High Bandwidth Memory~\cite{jedec:high} or Micron's Hybrid Memory Cube~\cite{hmc}. Each memory chip consists of multiple (e.g., 16-32) vertical DRAM partitions, called vaults, each with its own DRAM controller and signals. The near-memory processing units (MPUs) are scattered across the vaults as in prior work~\cite{ahn:scalable, gao:practical, pugsley:ndc}, while a network-on-chip (NoC) connects all the vaults to each other and to the off-chip links.



MPU-capable architectures consist of a pool of CPUs and memory chips. Fig.s~\ref{fig:overview}b,~\ref{fig:overview}c,~and~\ref{fig:overview}d show the memory organizations considered in this paper. The CPU is connected to multiple memory chips using high-speed point-to-point SerDes links and a packet-based communication protocol. Fig.~\ref{fig:overview}b depicts a star topology where the CPU is connected to a small number of memory chips~\cite{fujitsu:while, reinders:knights}. Larger memory systems interconnect dozens of chips in a daisy chain (Fig.~\ref{fig:overview}c), which minimizes the number of links~\cite{gao:practical, pugsley:ndc}, or a mesh (Fig.~\ref{fig:overview}d), which minimizes the number of hops~\cite{ahn:scalable, kim:memory-centric}. 


\begin{figure}
	\centering
	\includegraphics[width=0.9\columnwidth]{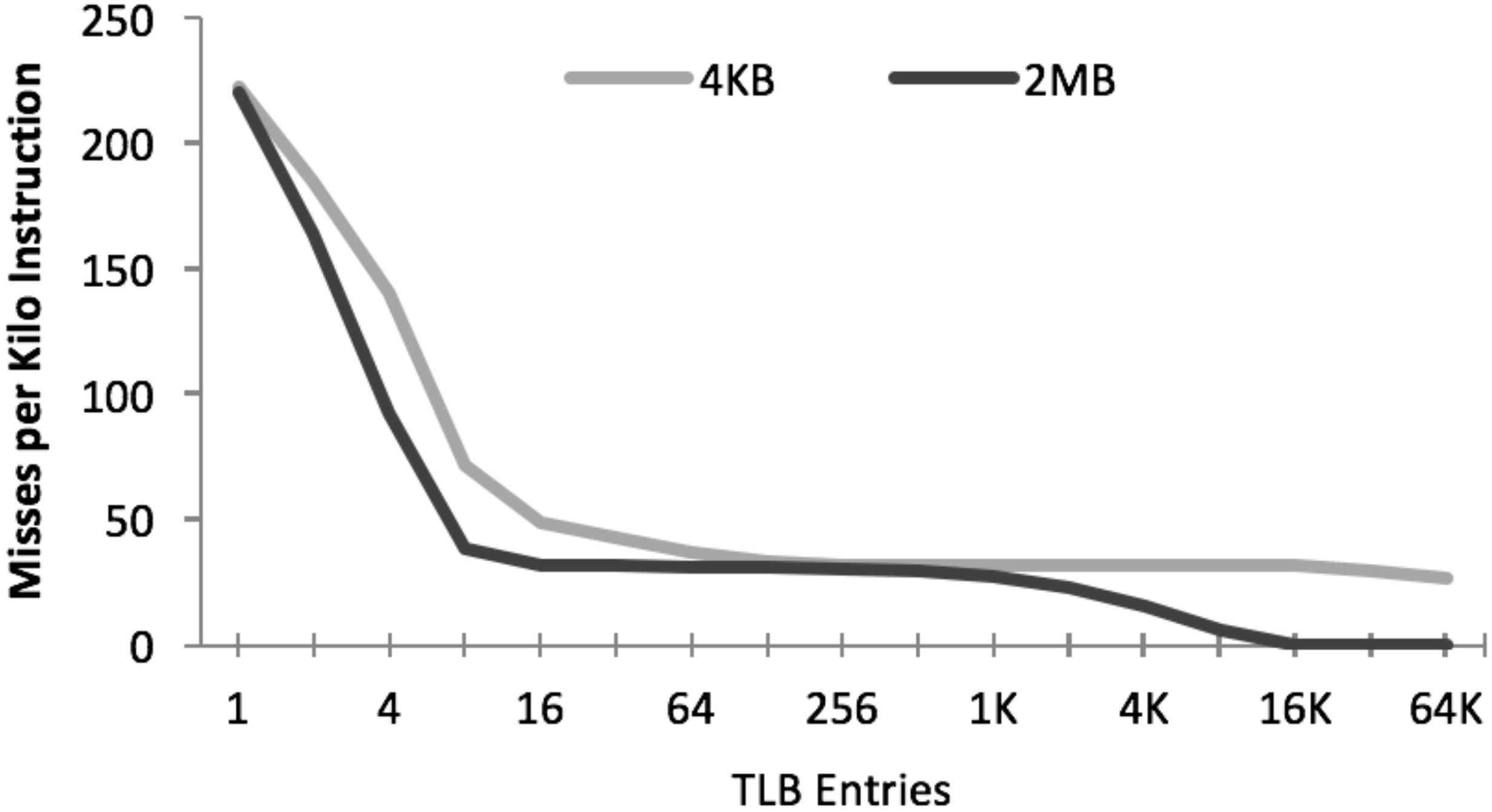}
	\caption{TLB miss rate as a function of TLB capacity.}
	\label{fig:tlbreach}
\end{figure}

\subsection{Unified Virtual Memory}
    Many applications could benefit from near-memory processing~\cite{gutierrez:integrated, ho:efficient, kocberber:meet, lim:thin, parashar:triggered, wu:navigating}. The widespread adoption of MPUs, however, depends on the efficiency and usability of their programming model. To enhance programmability, industry is moving towards unified virtual memory between CPUs and any computation unit in the system~\cite{nvidia:uva, rogers:amd}. Unified virtual memory enables ``pointer-is-a-pointer'' semantics~\cite{power:supporting, vesely:observation}, thus avoiding explicit and expensive data copies. More importantly, it provides a flat address space that is familiar to common programmers, while enforcing the required protection mechanisms to prevent compromising the security of the system.


To the best of our knowledge, no prior proposals for unified virtual memory between CPUs and MPUs are both general and efficient. Simple approaches would let either the CPU cores or an IOMMU translate addresses on behalf of MPUs~\cite{gao:practical, oskin:active, vesely:observation, xi:beyond}. These approaches incur a translation overhead of hundreds of nanoseconds as a recent study has shown~\cite{vesely:observation}. Unfortunately, applications with poor data locality, such as pointer chasing, would suffer from frequent CPU-MPU communication. Furthermore, providing conventional translation for MPUs implies high overhead as contemporary memories are beyond the reach of today's TLBs and MMU caches~\cite{basu:efficient,karakostas:redundant, pham:increasing, pham:colt}.

Figure~\ref{fig:tlbreach} compares the TLB miss rate for hash-table probes over a 32GB working set for 4KB and 2MB pages. The figure indicates that even with large pages and a TLB of 1K entries, for every thousand instructions, there are $40$ TLB misses, each requiring a page table walk.\footnote{The methodology is described in detail in Section~\ref{sec:evaluation}.} Walking the page table can be particularly expensive as it requires traversing multiple memory network hops, resulting in a TLB miss penalty that can exceed $500$ns. 


Such dramatic latency overhead can even hurt applications that perfectly partition data so that every MPU accesses data within its local memory~\cite{ahn:pim-enabled, ferdman:clearing, oskin:active, pugsley:ndc, xi:beyond}.
 Even in such cases, as page table entries are arbitrarily distributed across memory chips, the TLB miss penalty grows with the average network distance and quickly becomes the bottleneck, accounting for up to 70\% of the execution time. Note that this overhead is much larger for MPUs than it is for CPUs, where page table accesses can account for up to 50\% of execution time~\cite{basu:efficient, bhattacharjee:large-reach, karakostas:performance}. The reason is that the average distance from an MPU to other memory chips is significantly higher than the average CPU-memory distance for typical topologies (Figure~\ref{fig:overview} (b), (c) and (d)).


\begin{figure}
	\centering
	\includegraphics[width=\columnwidth]{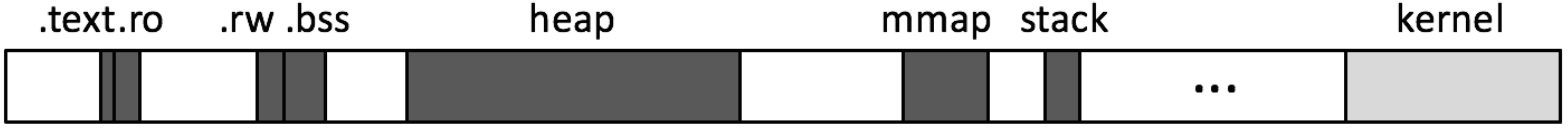}
	\caption{Typical virtual address space of a Linux Process. Dark-colored segments are exposed to the MPUs.}
	\label{fig:scope}
\end{figure}

Figure~\ref{fig:scope} shows the virtual address space layout of a Linux process~\cite{mauerer:professional}, featuring six virtual segment groups: the read-only segments, which consist of the binaries (.text) and globally visible constants (.ro); the read-write segments, containing global variables (.rw and .bss); the heap segment, which holds dynamically allocated objects; the mmap segments, for objects allocated through the \textit{mmap} syscall; the stack segment; and the kernel address space. We assume only the dark-colored segments are visible to MPUs. The virtual address space that is not exposed to MPUs (e.g., the kernel space) still enjoys full associativity.  



\section{Revisiting Associativity}
\label{sec:vm}

Page-based VM is an essential part of computer systems. At the time of its invention, the memory requirements of all active processes in the system exceeded the amount of available DRAM by orders of magnitude. A page table, which is a \textit{fully associative} software structure, was employed to maximize allocation flexibility by allowing any virtual page to map to any available \textit{page frame}. Interestingly, this architecture has barely changed and has only incorporated hardware structures to cache page table entries like TLBs~\cite{couleur:shared-access} and multi-level MMU caches~\cite{barr:translation, bhattacharjee:large-reach}.

With storage devices in recent decades dramatically lagging behind processors and memory in performance, and DRAM continuously improving in density and cost, many online services and analytic engines are carefully engineered to fit their working set in memory~\cite{oracle:timesten, sap:hana, basu:efficient, bronson:tao, ferdman:clearing, haque:few-to-many, kozyrakis:server, ousterhout:case, power:implications, zaharia:spark}. Due to rare page swapping, contiguous virtual pages are often mapped to contiguous physical pages~\cite{pham:increasing, pham:colt}, and hence the conventional page placement flexibility provided by full associativity remains largely unused. As such, we believe that the traditional \textit{full associativity} of VM (i.e., the flexibility to map a virtual page to any page frame) should be revisited. 


\begin{table}[t!]
 \begin{center}
  \caption{Workload description.}
  \scalebox{0.7}
  \small
  \vspace{0.01in}
  \label{table:workload}
  \renewcommand{\arraystretch}{1.0}
   {\scriptsize
    \begin{tabular}{ l  l }
     \toprule
      {\bf Workload}                  & {\bf Description}  \\
     	\toprule
      	\multirow{1}{*}{Cassandra}                       &  NoSQL data store running Yahoo's YCSB. \\
     	\cmidrule{2-2}
      	\multirow{1}{*}{Memcached}                      & Cache store running Twitter-like workload~\cite{lim:thin}. \\
	\cmidrule{2-2}
		\multirow{1}{*}{TPC-H}	& TPC-H on MonetDB column store (Q1-Q21). \\
	\cmidrule{2-2}
		\multirow{1}{*}{TPC-DS}	& TPC-DS on MonetDB column store (Queries of~\cite{kocberber:meet}). \\
		\cmidrule{2-2}
	 	\multirow{1}{*}{MySQL} 			& SQL storage engine running Facebook's LinkBench~\cite{facebook:linkbench}. \\
	\cmidrule{2-2}
		\multirow{1}{*}{Neo4j} 	&	Graph store running a neighbor traversal operation~\cite{musicbrainz}. \\
     \cmidrule{2-2}
      \multirow{1}{*}{RocksDB}                             &  Embedded store running Facebook benchmarks~\cite{facebook:rocksdb}. \\ 

     \bottomrule
    \end{tabular}
   } 
 \end{center}
  \vspace{-0.1in}
\end{table}

\sisetup{round-precision=2,round-mode=figures,scientific-notation=true}%

\begin{table*}[t]
	\centering
	\tiny
	\caption{ Impact of associativity on page conflict rate and page conflict overhead. }
	\resizebox{\linewidth}{!}{%
		\begin{tabular}{l!{\vrule width 0.5pt}ccccc!{\vrule width 0.5pt}ccccc}
			\toprule
			& \multicolumn{5}{c|}{Page conflict rate } & \multicolumn{5}{c}{Page conflict overhead norm. to memory latency}\\
			& \multicolumn{5}{c|}{(page conflicts per million accesses)} & \multicolumn{5}{c}{ (rate $\times$ penalty / memory latency)}\\
			& DM & 2-Way & 4-Way & 8-Way & 16-Way & DM & 2-Way & 4-Way & 8-Way & 16-Way \\
			\midrule
			RocksDB & \num{0.0842} & \num{0.0295} & \num{0.0250} & \num{0.0218} & \num{0} & 2.8\% & 0.98\% & 0.83\% & 0.73\% & \num{0} \\
			TPC-H & \num{1.0103} & \num{0.1586} & \num{0.0017} & \num{0} & --- & 33.68\% & 5.2\% & 0.06\% & \num{0} & --- \\
			TPC-DS & \num{0.1413} & \num{0.0002} & \num{0} & --- & --- & 4.71\% & 0.01\% & \num{0} & --- & --- \\
			Cassandra & \num{1.1166} & \num{0.0366} & \num{0.0003} & \num{0} & --- & 37.2\% & 1.22\% & 0.01\% & \num{0} & --- \\
			Neo4j & \num{39.072363} & \num{0.027753} & \num{0} & --- & --- & 1300.8\% & 0.93\% & \num{0} & --- & --- \\
			MySQL & \num{2.4145} & \num{0.0017} & \num{0} & --- & --- & 80.48\% & 0.06\% & \num{0} & --- & --- \\
			Memcached & \num{0.0496} & \num{0.0078} & \num{0} & --- & --- & 1.65\% & 0.26\% & \num{0} & --- & --- \\
			\bottomrule
		\end{tabular}%
	}
	\label{tab:rate_overhead}%
\end{table*}%

\sisetup{scientific-notation = false}%

Limiting associativity means that a page cannot reside anywhere in the physical memory but only in a fixed number of locations. For instance, direct-mapped VM maps each virtual page to a single page frame. Note that multiple virtual pages could map to the same physical frame, resulting in \emph{page conflicts}. Increasing the associativity adds more flexibility to the page mapping and reduces conflicts. To understand the associativity requirements, we collect long memory traces of applications (Table~\ref{table:workload}) that benefit from near-memory processing~\cite{ahn:scalable, gao:practical, gutierrez:integrated, kocberber:meet, lim:thin} using \textit{Pin}~\cite{luk:pin}. We extract the virtual page number (VPN) of each memory reference and use it to probe a set-associative structure. We then vary the associativity to study the conflicts.

\sisetup{scientific-notation = false}%

Table~\ref{tab:rate_overhead} (left) shows the page conflict rate as associativity varies. As shown, little associativity is enough to eliminate all page conflicts and match fully associative VM. On the one hand, Memcached and RocksDB do not exhibit frequent conflicts due to great contiguity in their virtual address space, as subsequent virtual pages are mapped to subsequent sets, never causing conflicts within a segment. On the other hand, Neo4j and Cassandra exhibit a large number of conflicts for a direct-mapped configuration because of their numerous randomly placed JVM segments which conflict with each other. However, conflicts drop fast, and 4 and 8 ways eliminate all the page conflicts for Neo4j and Cassandra respectively. The reason for page conflicts is two-fold: (i) the virtual space is not fully contiguous, and (ii) the software is unaware of the set-associative organization. Fortunately, the virtual space exhibits enough contiguity so that even unmodified software tolerates limited associativity.

Table~\ref{tab:rate_overhead} (right) estimates the average memory access time (AMAT) increase due to page conflicts. Here we conservatively assume that MPU's DRAM accesses are always local (the lower the memory latency, the higher the relative overhead of page conflicts). We also conservatively assume that page conflicts always generate a page fault to an HDD, taking $10$ms~\cite{symantec:getting}. Overall, limiting the VM associativity to $4$ ways introduces virtually zero overhead (e.g., adding less than $<1$\% to the AMAT in the worst case). This overhead and the required associativity would further decrease in the presence of faster SSD storage or a small fully associative software victim cache (as proposed before in the context of direct-mapped hardware caches~\cite{jouppi:improving}).

\section{DIPTA}
\label{sec:design}

To remove address translation completely from the critical path, it is \textit{necessary} and \textit{sufficient} to ensure that translation never takes more time than data fetch. Conventional translation hardware does not meet this requirement, as TLB misses---page walks---can take an unpredictable amount of time to resolve.

We exploit limited associativity to design a novel and efficient near-memory address translation mechanism. We propose DIPTA (Distributed Inverted Page Table), an address translation mechanism that completely eliminates the overhead of page walks. DIPTA restricts the associativity so that a page can only reside in a few number of physical locations which are physically adjacent--—i.e., in the same memory chip and DRAM row. Hence, all but a few bits remain invariant across the virtual-to-physical mapping, and with a highly accurate way predictor, the unknown bits are figured out so that address translation and data fetch are completely independent. Furthermore, to ensure that the data fetch and translation are completely overlapped, we place the page table entries next to the data in the form of an inverted page table, either in SRAM or embedded in DRAM. Hence, DIPTA completely eliminates the overhead of page walks. 

We first present a simple SRAM-based implementation of DIPTA and then present a scalable implementation where the translation information is embedded in DRAM.



\subsection{SRAM-based DIPTA}



As an effective way to ensure that the translation time never exceeds the data fetch time, we propose to distribute the translation information and co-locate it with the data, fetching them together to avoid exposing the translation latency. In the proposed architecture each DRAM vault keeps the information about the virtual pages it contains in an inverted page table. The resulting distributed page table is looked up in parallel with the data fetch.




The inverted page table (per vault) is implemented as a cache-like SRAM structure which is either direct-mapped or set-associative, depending on the associativity of VM. Assuming a 2GB MPU chip and $4$KB pages, DIPTA would contain 512K entries; one entry per page frame. Each entry holds the VPN of the page residing in the corresponding frame (36 bits) and the rest of the metadata, including 12 bits for the address space identifier (ASID) and 12 bits for page flags, totaling less than 8B for 48-bit virtual addresses (e.g., x86\_64, ARMv8).\footnote{Note that memory requests coming from MPUs contain both the VPN and ASID bits.} The page table would occupy $4$MB per MPU chip. Assuming 16-32 vaults~\cite{hmc,diram}, the per-vault SRAM overhead totals 128KB-256KB. For illustration purposes we assume $4$KB DRAM rows.


A direct-mapped implementation is trivial as the DIPTA SRAM lookup can proceed in parallel with the data fetch. The reason is that the virtual address enables direct indexing of both DIPTA and DRAM, as the virtual address uniquely identifies the DRAM row and column of the target cache block. The data fetch and translation can be issued independently and in parallel, the former to the vault's DRAM and the latter to its DIPTA partition. As the per-vault SRAM structure is small, fetching the translation is always faster than fetching the data from DRAM. By the time the cache block arrives, the translation metadata has already been fetched, and checked against the memory request, taking translation off the critical path.

\begin{figure*}[t]
	\centering
	\subfloat[Base layout]{
		\label{fig:base_layout}
		\includegraphics[width=0.70\columnwidth,clip]{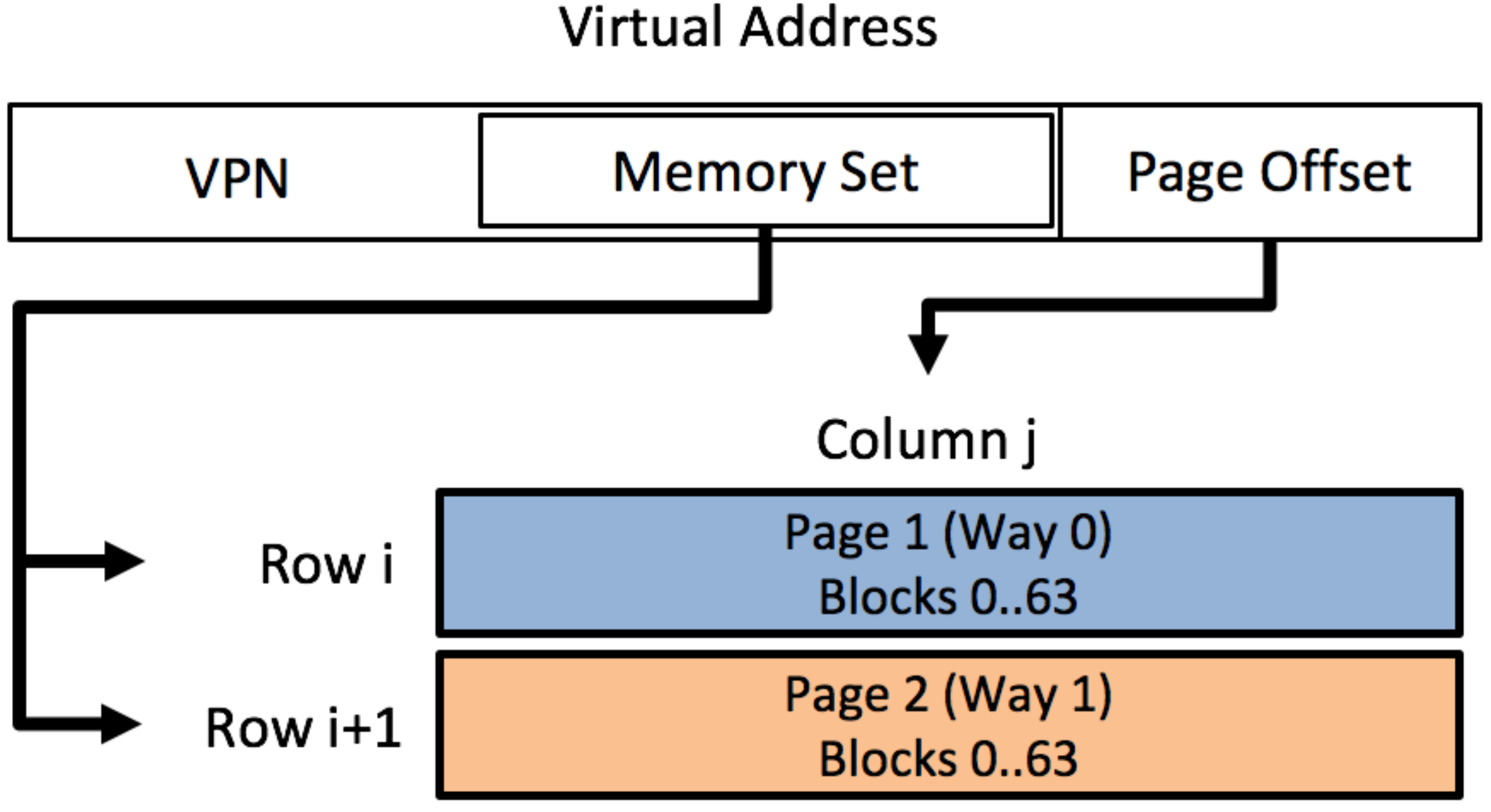}}
	\subfloat[DIPTA layout]{
		\label{fig:dipta_layout}
		\includegraphics[width=0.70\columnwidth,clip]{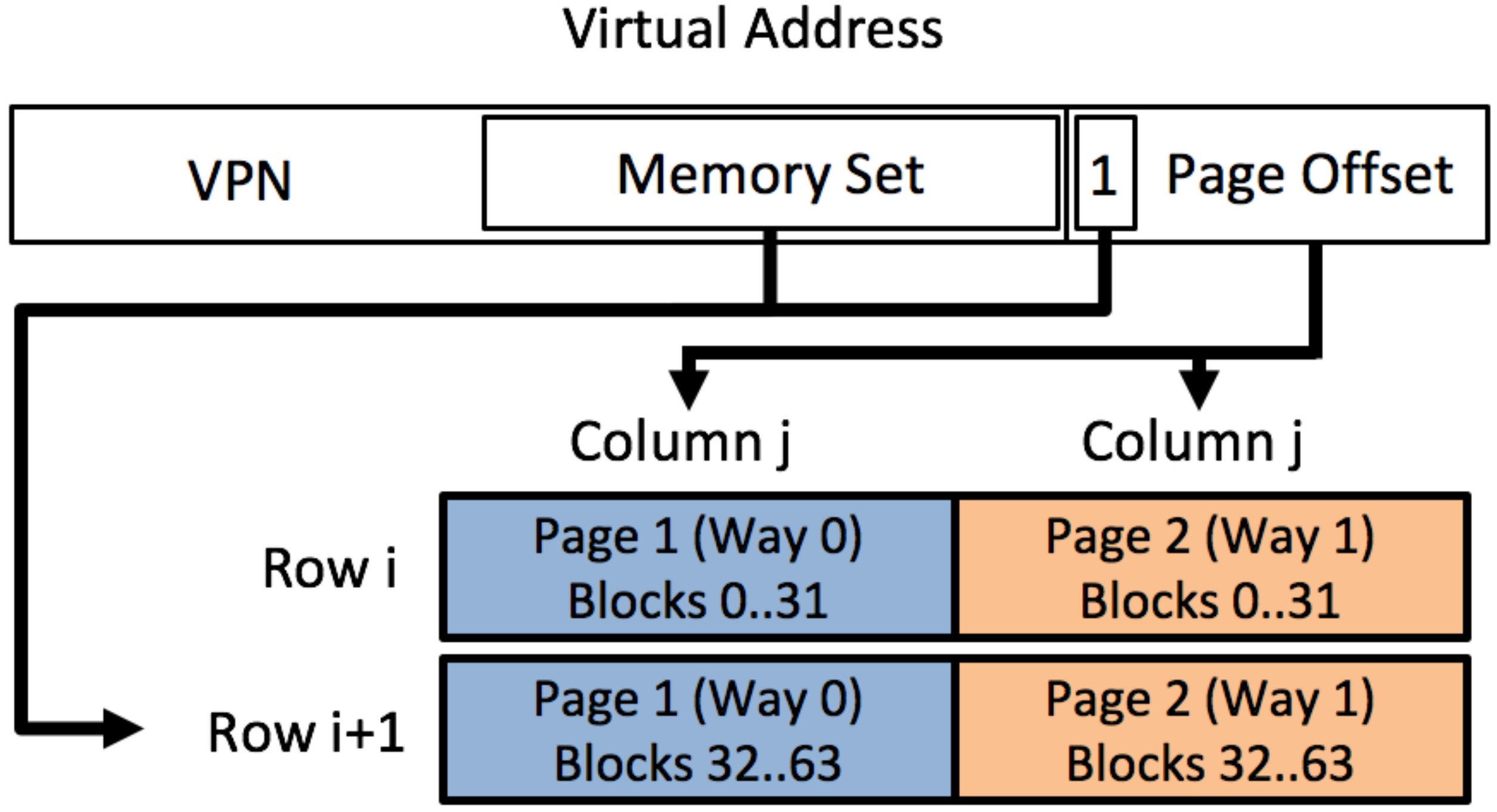}}
	\caption{Page layouts of two consecutive DRAM rows.
		\label{fig:placement}}
\end{figure*}



 
Supporting associativity is not trivial because a memory set now spans more than one page (depending on the associativity), and hence more than one DRAM row. As shown in Figure~\ref{fig:base_layout}, a virtual address can only identify a memory set uniquely, and hence the actual way (or DRAM row) where the target page resides is not known a priori. To determine which of the DRAM rows the target page resides, a lookup to the set-associative DIPTA partition is required. After the lookup, the DRAM row and column are used to fetch the data. Unfortunately, while such a solution is simple, the serial DIPTA lookup puts the translation back on the critical path.

 

We address the set-associative DIPTA lookup bottleneck in two ways. First, to locate the DRAM row where the target cache block resides, we propose to interleave pages within DRAM rows so that a virtual address uniquely identifies the DRAM row where the target cache block resides. As an example, Figure~\ref{fig:dipta_layout} illustrates the data placement of two 4KB pages into two 4KB DRAM rows for a 2-way set-associative VM. Because a DRAM row of 4KB cannot store two 4KB pages, we strip the data across DRAM rows by splitting each page into two parts. Even rows store the first half of each page, whereas odd rows store the second half. The target DRAM row is determined by the highest order bits of the part of the virtual address that used to identify the DRAM column (i.e., the page offset). In this example, it is the highest order bit as it breaks the page in two halves, which selects the second row (i.e., row i+1). This example can be easily generalized to any associativity.

Second, while interleaving pages allows for locating the target DRAM row, each row now contains chunks of multiples pages, and hence the bits of the virtual address that used to identify the DRAM column (i.e., the page offset) cannot uniquely determine which page and hence cache block to fetch, as shown in Figure~\ref{fig:dipta_layout}. A naive solution would read all the ways from the DRAM row at once and in parallel with translation, which would waste bandwidth and energy proportional to the associativity~\cite{jevdjic:unison,qureshi:fundamental}. To avoid such overheads, we employ a lightweight but highly accurate way prediction. Way prediction eliminates the need to fetch all the ways, while ensuring translation and data fetch happen independently, yet in a single DRAM access.






We design an address-based way predictor as they have been shown to achieve high accuracy for pages~\cite{calder:predictive, jevdjic:unison, powell:reducing}. Our way predictor is organized as a tagless array of $2^k$ entries indexed by a $k$-bit XOR hash of those VPN bits that determine the memory set within the vault (i.e., the bits that determine the vault are excluded). In the case of a 4-way associative system with $2$GB $16$-vault chips and $4$KB page frames, there are 13 bits that determine the set within each vault. For a way predictor of $32$ entries, we construct a $5$-bit XOR hash ($k$=$5$) for indexing. Each entry encodes the last accessed way in the set with two bits. In this case, the total storage for way prediction is only $8$B per vault, and covers $32$ sets or $128$ local pages in each vault. Because way prediction is done during the DRAM row activation, it is off the critical path. 


The combination of page interleaving and way prediction allows for fully overlapping the translation time in the common case of a predictor hit. Interleaving also minimizes the misprediction penalty; a second column access to an already opened DRAM row. Moreover, with the distributed nature of the predictor, the prediction accuracy in one vault is not affected by accesses to other vaults, boosting spatial and temporal locality.



\subsection{In-DRAM DIPTA} 

The SRAM solution is simple, but requires 4MB (16MB) of SRAM for a 2GB (8GB) memory stack. The area overhead grows linearly with the chip's capacity, leaving less space for MPUs. Dedicating a small fraction of a vault's DRAM to store DIPTA could completely eliminate the SRAM overhead. However, arbitrarily storing DIPTA in DRAM can make DIPTA lookups unpredictably long and difficult to overlap with the data fetch. Moreover, DIPTA lookups would contend for bandwidth with data fetches. 


A recent die-stacked DRAM cache proposal~\cite{jevdjic:unison} solves the tag-data co-location problem by dedicating the first 64-byte block in each DRAM row to store the tag metadata for the page residing in that row. Due to the limited associativity, the data location is independent of the tag content, hence the tag and read can be overlapped on cache hits~\cite{jevdjic:unison,qureshi:fundamental} by pipelining two separate back-to-back DRAM column accesses, one for the tag and the other for the data. Unfortunately, such a solution has an impact on the page size. Assuming a 4KB DRAM row, reserving 64B for metadata leaves $4096B-64B=4032B$ for the data, which may be acceptable for hardware DRAM caches, but it is certainly not acceptable for OS pages. 

\begin{figure}[t]
	\centering
	\includegraphics[width=0.85\columnwidth]{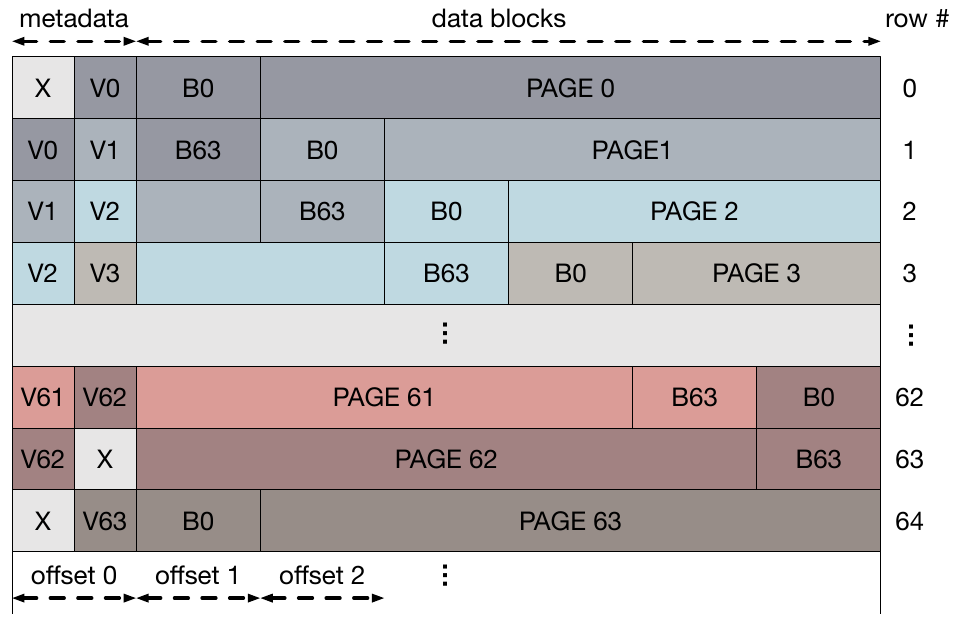}
	\caption{Metadata integration with 4KB pages \& rows.}
	\label{fig:layout}
\end{figure}

We instead propose a novel DRAM data layout illustrated in Figure~\ref{fig:layout}, where 63 4KB page frames are stored in 64 consecutive DRAM rows. For simplicity, we assume 4KB DRAM rows, although the solution can be trivially generalized to any page frame and DRAM row size. Because the first 64B in each row are reserved for metadata for all pages residing in that row, the last block of the first page cannot fit and is stored in the next row. In this example, each page spans exactly two DRAM rows, and each DRAM row contains blocks from at most two different pages. 

In Figure~\ref{fig:layout}, the first block of \emph{Page 0} is placed in the second block of \emph{Row 0}; we denote its position as \emph{Offset 1}, while the first block (\emph{Offset 0}) holds the metadata. The first half of the metadata block contains metadata of the page that ends in the current row (i.e., the page that could not fit in the previous row), whereas the second half contains metadata of the page that starts in the current row. Because \emph{Row 0} contains only one page, the first part of the metadata block is empty (denoted as \emph{X}). The last block (\emph{B63}) of \emph{Page 0} also occupies the first available data slot (\emph{Offset 1}), but in the subsequent row (\emph{Row 1}). The first block of the next page, \emph{Page 1}, occupies the block at \emph{Offset 2} in \emph{Row 1}, whereas the last block occupies the same position in the subsequent row, and so on. \emph{Page 62} starts at the very end of \emph{Row 62}, and occupies the entire \emph{Row 63}. Because no page starts in \emph{Row 63}, the second metadata slot in this row is empty. The layout of \emph{Row 64} is identical to the layout of \emph{Row 0}; rows 0-63 form a cycle. This solution incurs no SRAM overhead and requires dedicating 64B per DRAM row for metadata. The DRAM overhead is  $data\_line\_size/DRAM\_row\_size$ and decreases with the DRAM row size. For a 4KB DRAM row the overhead is 1/64 or $\sim$1.5\% of DRAM capacity. 

The target DRAM row for a given address is computed with minimal logic as: $block\_address/(k-1)$, where $k$ is the number of 64-byte blocks per row, in this example 64. The exact position of the requested data block is given by $block\_address$ mod $(k-1) +1$. Both formulas can be computed by a fairly simple hardware unit~\cite{qureshi:fundamental}.


Much as in SRAM DIPTA, the set-associative DRAM implementation follows the layout in Figure~\ref{fig:dipta_layout}. The first block in each row contains the metadata belonging to all the ways. The drawback of supporting high associativity in the DRAM-based implementation is that the storage overhead grows with associativity. For example, dedicating one block for metadata in a 2-way associative organization leaves an odd number of blocks for the data to store two ways, wasting a block to ensure symmetry, resulting in a DRAM overhead of $associativity \times data\_line\_size/DRAM\_row\_size$. However, the unused blocks could be used for more functionality (e.g., to store coherence directory entries).

\section{Discussion}
\label{sec:system}


\indent\textbf{Page faults.} The system must handle page faults triggered by MPU memory accesses. 
We choose to interrupt the CPU to run a handler, as MPUs may not be capable of running the OS. Upon a page fault, DIPTA responds to the MPU, notifying it of the fault. The MPU then places a request in a memory-mapped queue indicating the faulting virtual address and the responsible MPU's ID, and interrupts the CPU. After the missing page is brought into the memory, the handler updates the affected DIPTA entry with the new translation information. Once the fault is serviced, the handler notifies the appropriate MPU, which resumes its execution and retries the faulting address. Such page fault processing is also employed in today's integrated GPUs~\cite{vesely:observation}.

\textbf{TLB shootdowns \& flushes.} Our solution to maintain DIPTA entries coherent upon TLB shootdown and flush operations is similar to those used in integrated GPUs~\cite{vesely:observation}: An OS driver monitors any changes on virtual address spaces shared with the MPUs, triggering update operations on DIPTA for the affected entries. Note that the inverted nature of DIPTA eliminates any global coherence activity in the memory network because updates to DIPTA are fully localized to a single entry in the affected vault. DIPTA entries include the address space identifier (ASID) bits which avoids flushing all the entries of a given address space.

\textbf{Memory oversubscription.} We focus on workloads whose working set fits (almost) entirely in memory. We believe this scenario is the common case as memory is growing exponentially cheaper and bigger~\cite{gandhi:range}, while modern online services and analytic engines consistently require low response times~\cite{oracle:timesten, sap:hana, basu:efficient, bronson:tao, ferdman:clearing, haque:few-to-many, kozyrakis:server, ousterhout:case, power:implications, zaharia:spark}. Nevertheless, prior work on multi-GB caches backed up with an order-of-magnitude larger DRAM memory, has shown little sensitivity to associativity~\cite{jevdjic:unison, qureshi:fundamental}.


\textbf{Multiprogramming.} DIPTA supports multiprogramming as is, but may exhibit lower performance due to contention for associativity. In the case of first-party workloads, servers are often dedicated to in-memory services which take up all resources to ensure performance isolation~\cite{mars:bubble-up, yang:bubble-flux}, or overprovisioned to operate at low to modest loads to achieve responsiveness~\cite{barroso:case, haque:few-to-many, jeon:predictive, mcmillan:data}. In virtualized environments, however, where many applications are consolidated on the same server, limited associativity may be an issue.  In this case, the OS could be aware of the associativity to properly choose virtual addresses during segment allocation to minimize the conflicts.

\textbf{Synonyms.} As with any inverted page table, synonyms are not straightforward to handle~\cite{jacob:virtual}. A trivial approach would enforce synonyms to either have the same virtual addresses or to map to the same set~\cite{cheng:virtual}, and extend each DIPTA entry with extra storage. A more clever approach, inspired by Yoon and Sohi's work~\cite{yoon:revisiting}, would add a small per-vault structure populated by the OS to remap synonym pages to a single leading virtual page, and consequently to a single page frame. In this work, we do not extend DIPTA to support synonyms because we do not expose shared libraries or the kernel address space, which are the sources of synonyms~\cite{basu:reducing, yoon:revisiting}.

\textbf{Cache hierarchy.} In case MPUs integrate physical caches, a naive approach would add a TLB to cache frequently used page table entries, while TLB misses would be accelerated by DIPTA. Upon a TLB miss, the translations and data would be accessed in parallel (as part of the normal operation), but cached in separate structures. A more natural design, which also avoids TLBs and TLB shootdowns~\cite{villavieja:didi}, is to use virtual caches. Recent practical designs for virtual cache hierarchies would be a perfect fit for DIPTA~\cite{park:efficient, yoon:revisiting}. In this approach, MPUs access the cache with virtual addresses, and upon a cache miss, the request is propagated to DIPTA to translate and fetch the corresponding block.   



\textbf{Multi-level memories} Although prior work on memory-side processing assumes a single level~\cite{ahn:scalable,  ahn:pim-enabled, gao:practical, pugsley:ndc}, memory can be organized as a hierarchy, with a die-stacked cache~\cite{reinders:knights, volos:fat} backed up by planar memory. For hardware-managed caches, DIPTA performs the translation and accesses the page frame speculatively, and in case the page frame is not in the cache, it is fetched from planar memory as part of the standard cache miss operation. Once the page frame is in the appropriate DRAM row, the data is sent back to the MPU. The DIPTA page table entries have to be embedded in both planar and die-stacked DRAM, and move with the page frame. In software-managed hierarchies~\cite{reinders:knights}, MPUs rely on the software API for explicit migration of pages into the die-stacked memories, as MPUs cannot access planar memory directly. As part of the page migration operation, the DIPTA page table entries are populated accordingly.

\textbf{Operating system support.} The operating system only needs to guarantee that the virtual page number and the page frame number map to the same memory set. OSs that support virtual caches already provide this capability (e.g., Solaris~\cite{cheng:virtual} and MIPS OS~\cite{taylor:tlb}).

\section{Evaluation}
\label{sec:evaluation}


\subsection{Methodology}

Like the recent work on virtual memory~\cite{barr:spectlb, basu:efficient, bhattacharjee:large-reach, papadopoulou:prediction-based, pham:increasing, pham:colt, saulsbury:recently-based}, we use a combination of trace-driven functional and full-system cycle-accurate simulation.


\subsubsection{Performance}



Full-system simulation for the server workloads listed in Table~\ref{table:workload} is not practical. Hence, we resort to the CPI models often used in VM research~\cite{bhattacharjee:shared, papadopoulou:prediction-based, saulsbury:recently-based} to sketch the performance gains. These prior studies report performance as the reduction in the translation-related cycles per instruction. As CPI components are additive, this metric is valid irrespective of the workload's baseline CPI. We further strengthen this methodology by studying the CPI savings on all memory cycles, not only on translation stalls (as we overlap translation and data fetch operations). Our model thus captures both the translation cycles and data fetch cycles, which together constitute the largest fraction of the total CPI in server workloads~\cite{ferdman:clearing}. Hence, our results are more representative of the end-to-end benefits of each technique. The CPI is measured by feeding the memory traces into our cycle-accurate simulator.


Furthermore, we evaluate a set of data-structure traversal kernels---ASCYLIB~\cite{david:asynchronized}---in full-system cycle-accurate simulation. ASCYLIB contains state-of-the-art multi-threaded hash tables, binary trees, and skip lists. For clarity, we present results for four representative implementations: Java Hash Table (Hash Table), Fraser Skip List (Skip List), Howley Binary Search Tree (BST Internal), and Natarajan Binary Search Tree (BST External). We choose this specific suite because dynamic data structures are the core of many server workloads (e.g., Memcached's hash table, RocksDB's skip list), and are a great match for near-memory processing~\cite{hsieh:accelerating, kocberber:meet}. The abundance of pointer chasing results in poor locality which allows us to stress the translation and way prediction mechanisms.



\begin{table}[b!]
	\begin{center}
		\caption{System parameters.}
		\scalebox{0.7}
		\small
		\vspace{0.01in}
		\label{table:dipta_system}
		\renewcommand{\arraystretch}{1.0} 
		{\scriptsize
			\begin{tabular}{ l  l }
				\toprule
				{\bf MPU logic}                  & {\bf Description}  \\
				\toprule
				\multirow{1}{*}{Cores}                       &  Single-issue, in-order, 2GHz \\ 
				\cmidrule{2-2}
				\multirow{1}{*}{L1-I/D}                       &  32KB, 2-way, 64B block, 2-cycle load-to-use \\ 
				\toprule
				{\bf MMU}                  & {\bf Description}  \\
				\toprule
				
				
				\multirow{3}{*}{TLB}	& $4$KB pages: 64-entry, 4-way associative \\
				& $2$MB pages: 32-entry, 4-way associative \\
				& $1$GB pages: 4-entry, fully associative \\
				\cmidrule{2-2}
				\multirow{1}{*}{STLB} 			& 4KB/2MB pages: 1024-entry, 8-way associative \\ 
				
				\cmidrule{2-2}
				\multirow{3}{*}{Caches} 	&	L4: 2-entry, fully associative~\cite{bhattacharjee:large-reach} \\ 
				&       L3: 4-entry, fully associative~\cite{bhattacharjee:large-reach} \\
				&       L2: 32-entry, 4-way associative~\cite{bhattacharjee:large-reach} \\
				
				\toprule
				{\bf Memory}                  & {\bf Description}  \\
				\toprule

				\multirow{1}{*}{MPU chip}                                   &  8GB chips, 8 DRAM layers x 16 vaults \\         
				\cmidrule{2-2}
				\multirow{1}{*}{Networks}                             &  4, 8, 12, and 16 chips in daisy chain and mesh \\       
				\cmidrule{2-2}
				\multirow{2}{*}{DRAM}                                &  $t_{CK} = 1.6$ns, $t_{RAS} = 22.4$ns, $t_{RCD} = 11.2$ns \\
				&  $t_{CAS} = 11.2$ns, $t_{WR} = 14.4$ns, $t_{RP} = 11.2$ns \\ 
				\cmidrule{2-2}
				\multirow{1}{*}{Serial links}                             &  2B bidirectional, 10GHz, 30ns per hop~\cite{kanter:cavium, towles:unifying} \\           
				\cmidrule{2-2}
				\multirow{1}{*}{NoC}                             &  Mesh, 128-bit links, 3 cycles per hop \\
				\toprule
				{\bf DIPTA}                  & {\bf Description}  \\
				\toprule
				\multirow{1}{*}{Configuration}                             &  4-way associative, 1024-entry WP \\

				\bottomrule
			\end{tabular}
		} 
	\end{center}
	\vspace{-0.1in}
\end{table}

\subsubsection{Workloads}
For the associativity experiments in Sections~\ref{sec:vm} and ~\ref{sec:proposals}, we collect long memory traces using \textit{Pin}~\cite{luk:pin}. For workloads with fine-grained requests (i.e., Memcached, RocksDB, MySQL, and Cassandra), the traces contain the same number of instructions as the application executes in 60 seconds without Pin. For analytics workloads (i.e., MonetDB and Neo4j), we instrument the entire execution. We feed the traces into a tool that models a set-associative memory of $8$GB, $16$GB, and $32$GB. For the associativity experiments of Section~\ref{sec:vm}, the workloads are tuned to use $8$GB of memory. Then, we scale the workloads up to $16$GB and $32$GB for Section~\ref{sec:proposals}'s experiments. 

The traces are collected on a dual-socket server CPU (Intel Xeon E5-2680 v3) with $256$GB of memory, using the Linux 3.10 kernel and Google's TCMalloc~\cite{google:tcmalloc}. Address space randomization (ASLR) is enabled in all experiments. 

For the performance experiments of Section~\ref{sec:evaluation}, we employ the server traces of 32GB and 64GB, depending on the size of the network. We use 32GB and 64GB for the 4-chip and 16-chip configurations respectively. For the data-structure kernels, each workload performs uniformly distributed key lookups on its in-memory data-structure. The datasets range from 16GB to 20GB (depending on the workload) across all network configurations.

\subsubsection{Simulation Parameters}
We use the Flexus full-system simulator~\cite{wenisch:simflex}, with detailed core, MMU, memory hierarchy, and interconnect models. Following prior work on near-memory processing, which assumes single-issue in-order cores~\cite{ahn:scalable, gao:practical, pugsley:ndc}, we model the MPU cores after ARM Cortex A7~\cite{arm:cortex-a7}. We privilege the baseline with a high-end MMU similar to Intel Xeon Haswell~\cite{intel:tlbs, mammarlund:4th}, with multi-level TLBs and MMU caches~\cite{barr:translation, bhattacharjee:large-reach}. We assume a 4-level hierarchical radix tree page table ~\cite{jacob:look} with 48-bit virtual and physical addresses (as in ARMv$8$ and x$86\_64$). The MMU supports $4$KB, $2$MB, and $1$GB pages. Note that page table entries are transparently allocated in the L$1$-D cache. We probe the cache with physical addresses for the baseline and with virtual addresses for DIPTA. We verify that TLB misses never reference a cache-resident block, and therefore virtual and physical caches behave identically.

We assume the Hybrid Memory Cube organization with eight $8$Gb DRAM layers and 16 vaults~\cite{micron:hmc}. We conservatively estimate the die-stacked memory timing parameters from publicly available information and research literature~\cite{gao:practical}. We employ a 4-way VM implementation of DIPTA with a way-predictor of 1024 entries per vault. The DRAM and SRAM implementations provide almost identical results, with the tradeoff being between SRAM area and DRAM capacity. The DRAM implementation has practically no SRAM overhead (except for tiny way predictors) but occupies space in DRAM for translations. The SRAM overhead for 8GB chips is 16MB (partitioned across vaults) for an area of $20mm^2$ in $22$nm, corresponding to only 9\% of the area of an $8$Gb DRAM die (e.g., $226mm^2$~\cite{shevgoor:quantifying}). Its access latency of 8 cycles guarantees that the memory and DIPTA accesses are overlapped. Table~\ref{table:dipta_system} shows the system parameters.

\subsection{Results}
We study the translation overheads as we vary the topology and scale of the memory network, as well as the amount of data locality with respect to the MPU chips.

\subsubsection{Way Prediction Accuracy}
\label{sec:wayprediction}
\begin{figure}
	\centering
	\includegraphics[width=0.85\columnwidth]{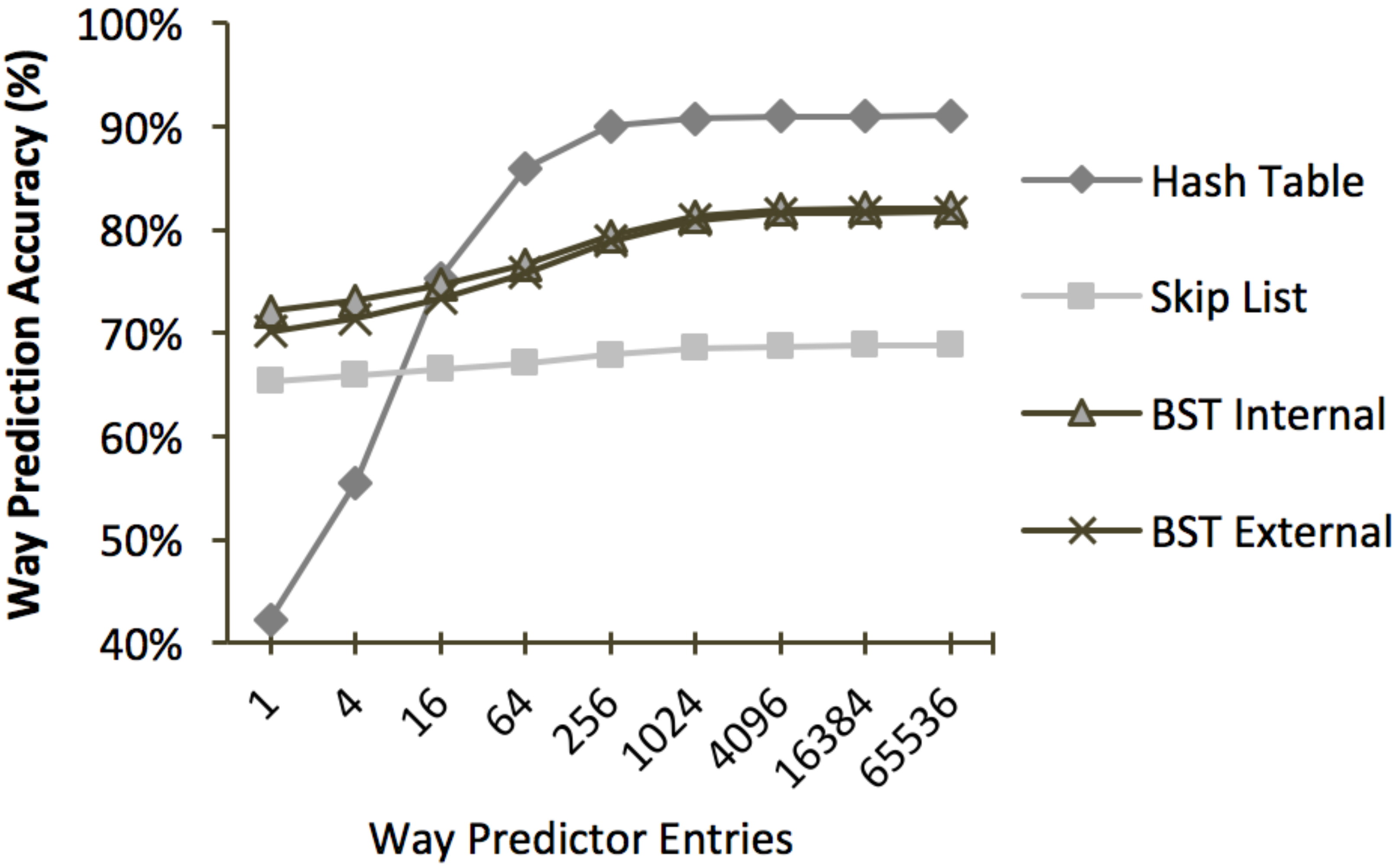}
	\caption{Way prediction accuracy.}
	\label{fig:wayprediction}
\end{figure}

The way prediction accuracy on our server workloads (Table~\ref{table:workload}) ranges from $96$\%-$99$\% due to spatial and temporal locality. To better stress the way predictor, we study its accuracy on ASCYLIB, which could be considered the worst case (e.g., the behavior of Skip Lists is very similar to GUPS). Figure~\ref{fig:wayprediction} shows the way prediction accuracy for a 4-way associative organization as the number of entries increases. A single entry requires $2$ bits (4-ways), yielding very high accuracy, $69$\%-$91$\%, at a tiny storage cost. Besides leveraging the spatial locality, the predictor's distributed nature also boosts the temporal locality: the accuracy in one vault is not affected by accesses to others. In this work, we assume a 1024-entry way predictor per vault, incurring only $256$B per vault and $4$KB per chip storage overhead. 


\begin{figure*}
	\centering
	\includegraphics[width=0.85\textwidth]{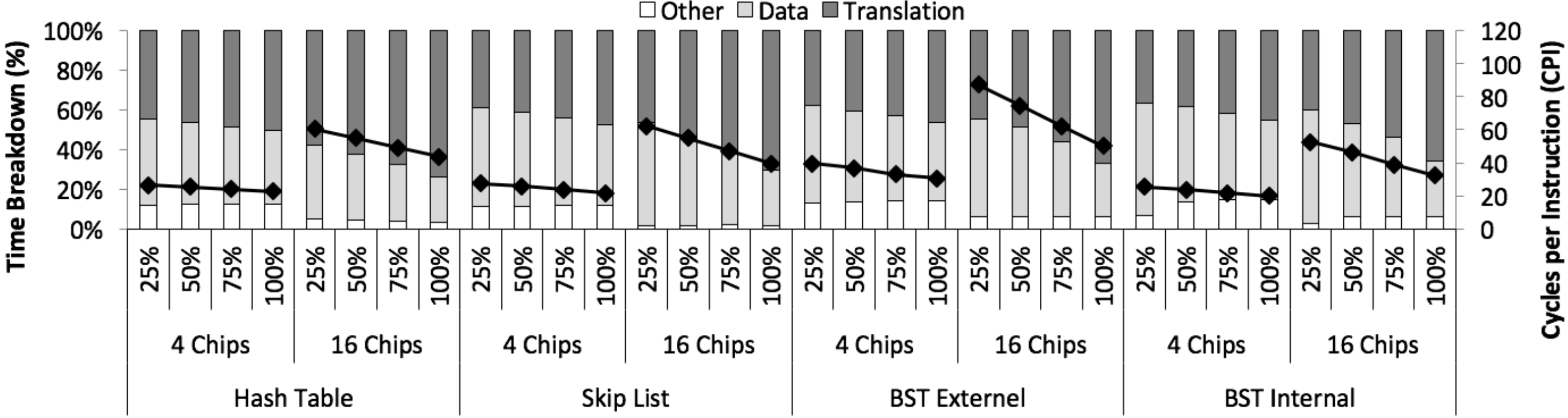}
	\caption{Time breakdown and CPI for different data locality and network topologies.}
	\label{fig:analysis}
\end{figure*}

\subsubsection{Where Does the Time Go?}
\label{sec:breakdown}

Figure~\ref{fig:analysis} shows the execution time breakdown and CPI of the data-structure kernels for the conventional translation with 4KB pages. We perform this experiment for 4- and 16-chip daisy chain topologies to see the impact of the network size. We also control the fraction of local data accesses, varying it from $25$\% to $100$\%---an application that perfectly partitions its dataset across the memory chips exhibits $100$\% data locality. As the figure shows, improving data locality reduces the overall execution time due to fewer cross-chip data accesses, but also increases the relative contribution of address translation to the total execution time---measuring up to 70\% with high locality.

The tree data structures kernels (i.e., BST Internal/External) show slightly better TLB locality compared to the Hash Table and Skip List kernels. This locality is exhibited in the top tree levels. Such locality is not present in the hash table and skip list data structures, where probes for different keys will likely access distinct pages. Additionally, given a data locality point, the translation overhead significantly increases with the network size for all the kernels, as expected. Overall, the data-structure kernels exhibit significant translation time, which usually increases with the fraction of data locality and memory chip count. Hence, reducing the cycles spent in translation has the potential to bring great performance benefits.


\subsubsection{Performance Analysis}
\label{sec:performance}

\begin{figure*}[t]
	\centering
	\subfloat[Hash Table]{
		\label{fig:speedup_ht}
		\includegraphics[width=0.4\textwidth,clip]{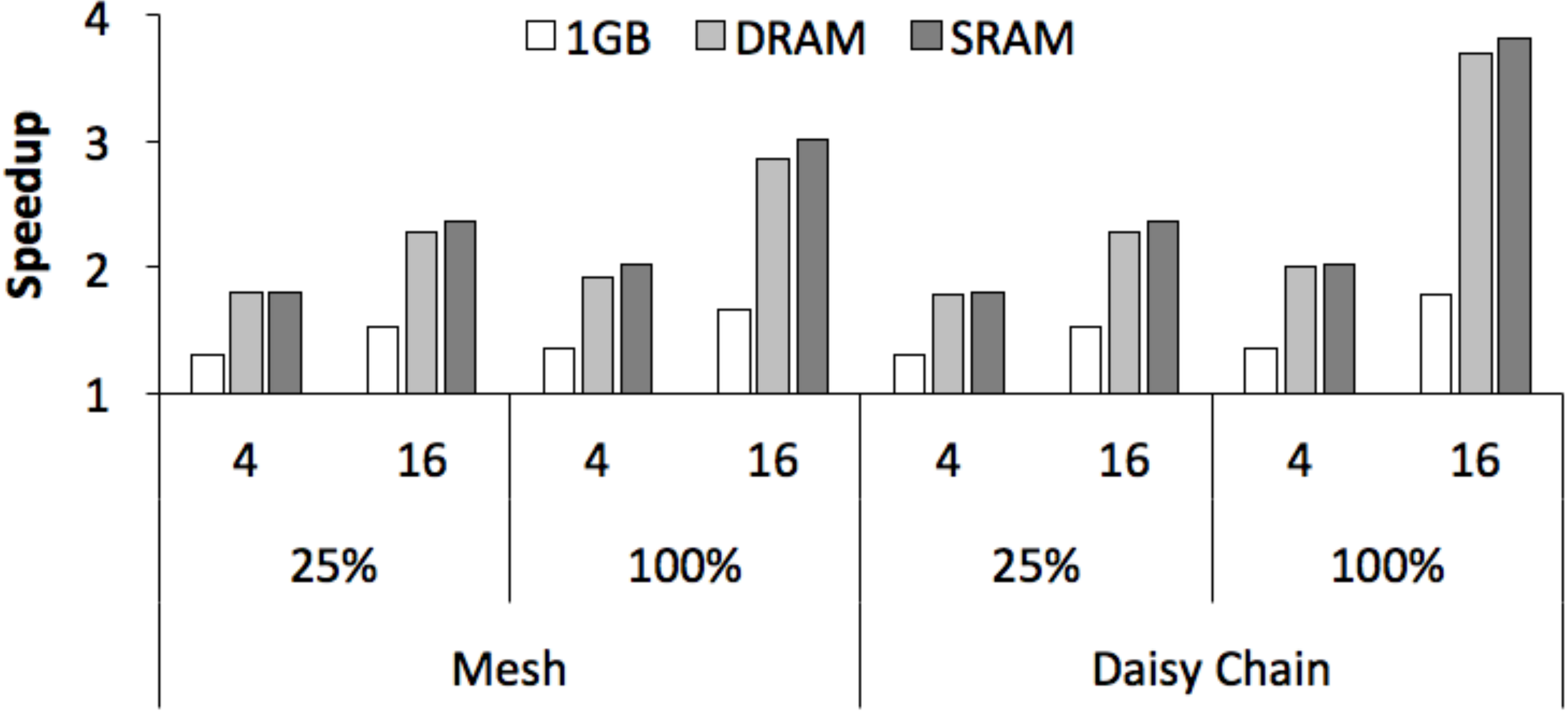}}
	\subfloat[Skip List]{
		\label{fig:speedup_sl}
		\includegraphics[width=0.4\textwidth,clip]{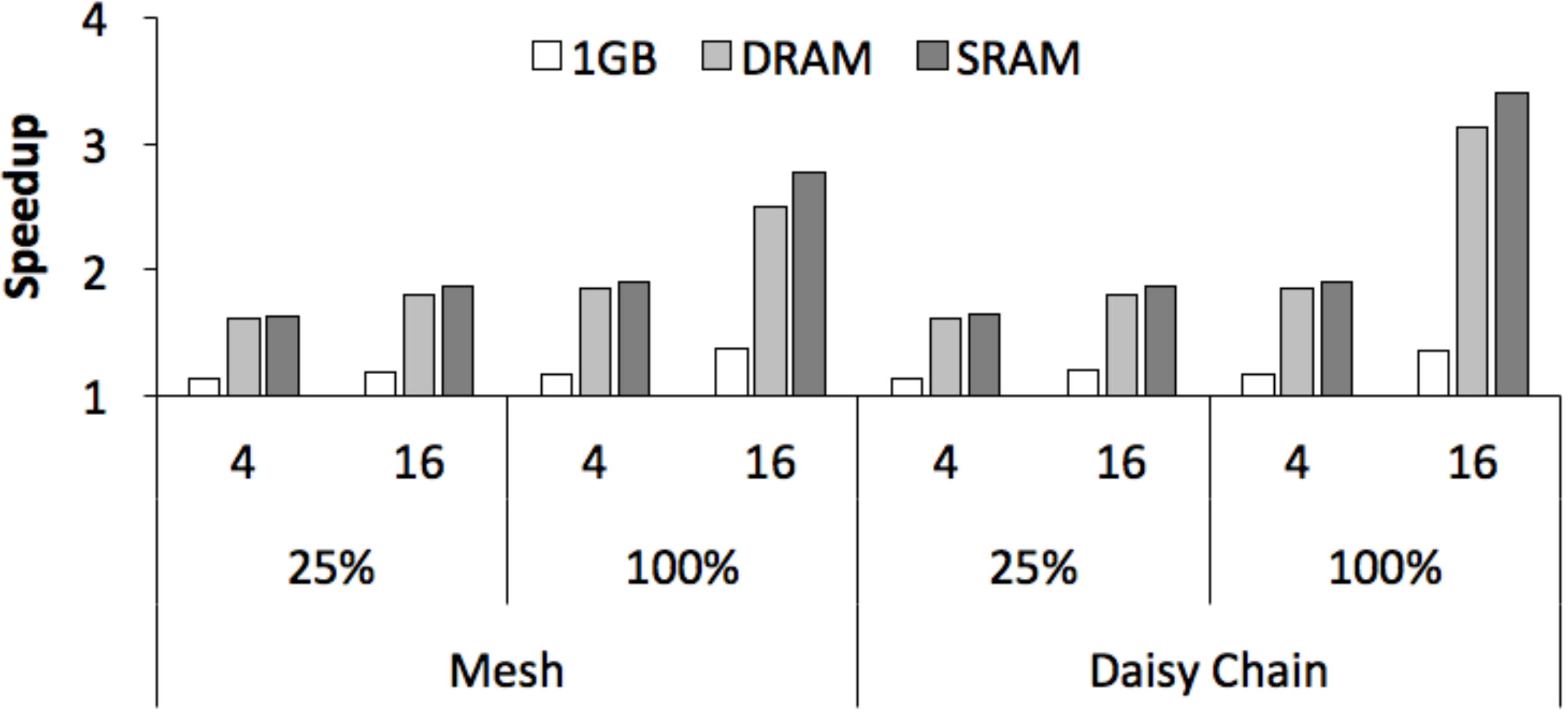}}
	\hspace{.01in}
	\subfloat[BST External]{
		\label{fig:speedup_bste}
		\includegraphics[width=0.4\textwidth,clip]{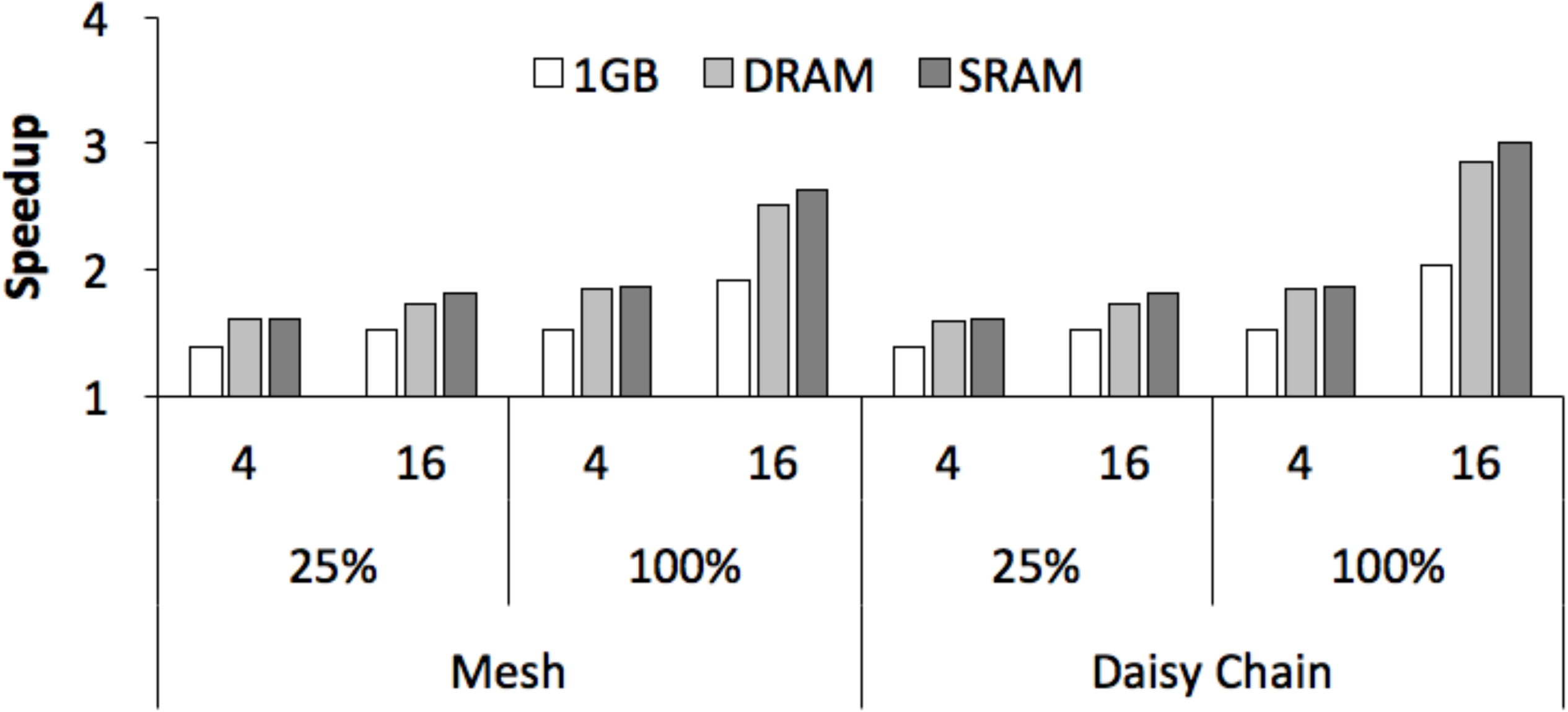}}
	\subfloat[BST Internal]{
		\label{fig:speedup_bsti}
		\includegraphics[width=0.4\textwidth,clip]{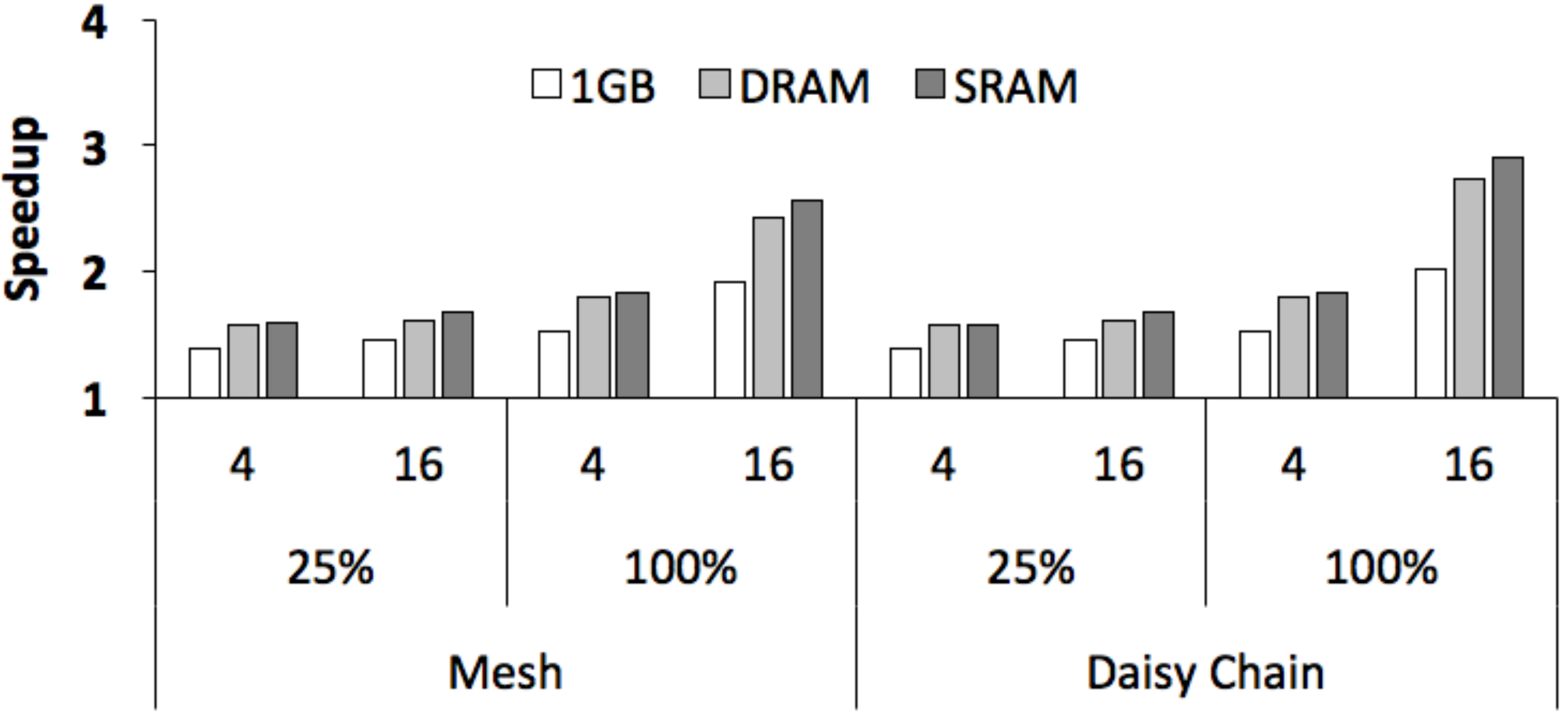}}
	\caption{Speedup results over 4KB pages for the data-structure kernels.
		\label{fig:speedup_micro}}
\end{figure*}

Figure~\ref{fig:speedup_micro} shows the speedup that the baseline with 1GB pages and both DRAM and SRAM DIPTA implementations provide over the baseline with 4KB pages, for 4- and 16-chip mesh and daisy chain topologies, for the data-structure kernels. For space reasons we present the results for the extreme locality points only: $25$\% and $100$\%; 
As expected, the speedup grows with locality as well as with the average distance in the memory network. Furthermore, the speedups on the daisy chain are more pronounced as the average hop count is larger than in the mesh. In the Hash Table and Skip List kernels, which exhibit the poorest TLB behavior, translation accounts for the largest fraction of the execution time among all benchmarks, which is reflected in the speedup. In contrast, the tree-based data structures exhibit better data and TLB locality, and consequently, lower speedups. SRAM DIPTA's speedups over 4KB pages range between $1.58\times$ and $3.81\times$, with an average speedup of $2.11\times$, whereas the baseline with 1GB pages improves the performance by $1.20\times$ to $2.02\times$, with an average speedup of $1.45\times$. SRAM DIPTA's speedups over 1GB pages range between $1.14\times$ and $2.13\times$, with an average of $1.44\times$. DRAM DIPTA performs within $4\%$ of SRAM DIPTA on average. Although omitted for brevity, we also compare DIPTA against the baseline with $2$MB pages, which performs only slightly better than 4KB pages in most cases. As shown in the results, DIPTA significantly outperforms conventional address translation hardware. Note that DIPTA virtually eliminates the overhead of address translation, and hence our results are equal to the ideal translation mechanism of a perfect TLB.

\begin{table*}[ht]
	\centering
	\tiny
	\caption{ Analysis of virtual segments as dataset scales }
	\resizebox{\textwidth}{!}{%
		\begin{tabular}{l!{\vrule width 0.5pt}ccc!{\vrule width 0.5pt}ccc!{\vrule width 0.5pt}ccc!{\vrule width 0.5pt}ccc}
			\toprule
			& \multicolumn{3}{c|}{Total segments} & \multicolumn{3}{c|}{99\% coverage} & \multicolumn{3}{c|}{Largest segment} & \multicolumn{3}{c}{Largest 32 segments}\\
			& 8GB & 16GB & 32GB & 8GB & 16GB & 32GB & 8GB & 16GB & 32GB & 8GB & 16GB & 32GB\\
			\midrule
			RocksDB & \num{209} & \num{370} & \num{692} & \num{160} & \num{323} & \num{638} & \num{0.62}\%  & \num{0.31}\% & \num{0.16}\% & \num{19.87}\%  & \num{9.94}\% & \num{4.97}\% \\
			TPC-H & \num{275} & \num{275} & \num{292} & \num{45} & \num{45} & \num{52} & \num{23.70}\% & \num{23.74}\% & \num{22.79}\% & \num{94.32}\% & \num{94.38}\% & \num{94.53}\% \\
			TPC-DS & \num{419} & \num{397} & \num{401} & \num{173} & \num{165} & \num{162} & \num{9}\% & \num{3.3}\% & \num{3.3}\% & \num{60.85}\% & \num{50.08}\% & \num{50.37}\% \\
			Cassandra & \num{392} & \num{410} & \num{523} & \num{25} & \num{33} & \num{132} & \num{59}\% & \num{33}\% & \num{20}\%  & \num{99.31}\% & \num{98.98}\% & \num{68.41}\% \\
			Neo4j & \num{204} & \num{889} & --- & \num{30} & \num{207} & --- & \num{59}\% & \num{39.97}\% & --- & \num{99.97}\% & \num{50.83}\% & --- \\
			MySQL & \num{145} & \num{145} & \num{145} & \num{2} & \num{2} & \num{2} & \num{96.6}\% & \num{96.80}\% & \num{97.22}\% & \num{99.97}\% & \num{99.99}\% & \num{99.99}\% \\
			Memcached & \num{52} & \num{52} & \num{52} & \num{1} & \num{1} & \num{1} & \num{99.97}\% & \num{99.98}\% & \num{99.99}\% & \num{100}\% & \num{100}\% & \num{100}\%\\
			\bottomrule
		\end{tabular}%
	}
	\label{tab:segments_table}%
\end{table*}%







\begin{figure}
	\centering
	\includegraphics[width=0.9\columnwidth]{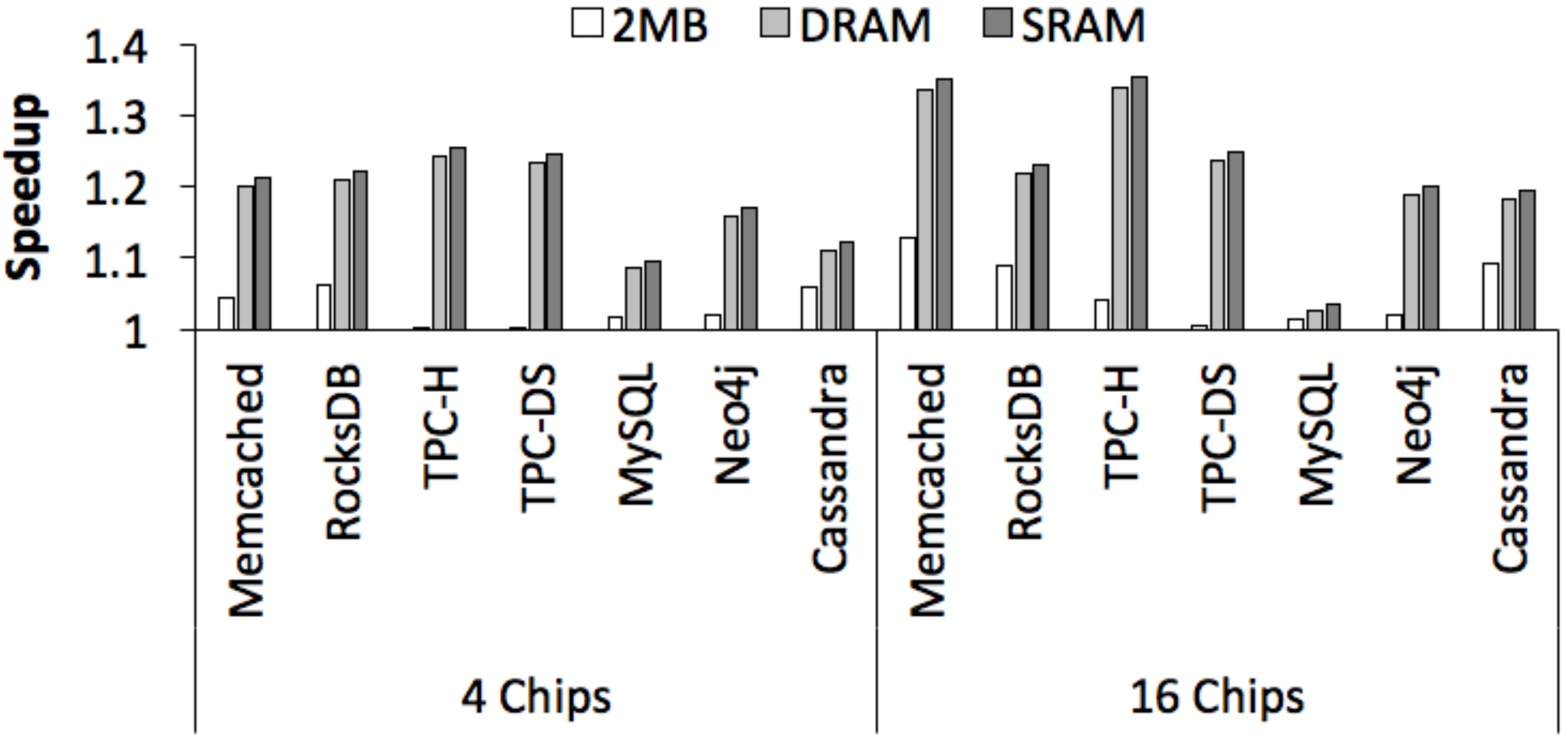}
	\caption{DIPTA and 2MB pages speedup over 4KB pages.}
	\label{fig:speedup_server}
\end{figure}

 Figure~\ref{fig:speedup_server} shows the speedup of DRAM and SRAM DIPTA implementations over the baseline that uses 4KB pages for the server workloads on a mesh topology. The figure also shows the impact of using 2MB pages. For the 4-chip configuration, the speedup of SRAM DIPTA over 4KB pages ranges between $1.09\times$ and $1.25\times$, with an average of $1.19\times$. Additionally, SRAM DIPTA's speedup over the baseline with 2MB pages ranges between $1.07\times$ and $1.25\times$, with an average of $1.15\times$. For the 16-chip configuration, the speedup of SRAM DIPTA over 4KB pages ranges between $1.03\times$ and $1.35\times$, with an average of $1.23\times$. Lastly, DIPTA's speedup over 2MB pages ranges between $1.02\times$ and $1.29\times$, with an average of $1.16\times$. DRAM DIPTA performs within $1\%$ of SRAM DIPTA on average. For clarity, we omit the results with 1GB pages, which performs better than 4KB pages, but always worse than 2MB pages. Employing 1GB pages performs worse than 2MB pages because the number of entries in the MMU for 1GB pages is significantly limited; there are only four entries. As for the data-structure kernels, DIPTA clearly outperforms conventional translation hardware, while virtually delivering the performance of an ideal translation with a perfect TLB. 

\subsubsection{Comparison with Other Proposals}
\label{sec:proposals}

Two recent proposals on address translation for CPUs are direct segments (DS)~\cite{basu:efficient} and redundant memory mappings (RMM)~\cite{gandhi:range}. These approaches exploit the abundant contiguity available in the virtual address space of certain applications by mapping one (for DS) or a few (for RMM) virtual segments to contiguous page frames. 


A comparison of DIPTA with DS and RMM on ASCYLIB is trivial, as these data-structure kernels have a very simple memory layout where all the data is mapped to a single virtual segment. To perform the more challenging comparison of these techniques on our set of server workloads, we analyze the maximum contiguity available in their virtual address space. We employ Linux's \textit{pmap} tool to periodically scan their memory structure. The results are presented in Table~\ref{tab:segments_table}. \textit{Total segments} represents the total number of virtual segments. \textit{99\% coverage} indicates the number of virtual segments required to cover 99\% of the physical address space. \textit{Largest segment} shows the fraction of the physical address space covered with the largest virtual segment. \textit{Largest 32 segments} shows the fraction of the physical space covered with the largest 32 segments. Note that these results represent an ideal case for DS and RMM. We employ datasets of $8$GB, $16$GB, and $32$GB. For Neo4j, we use two graphs of 8GB~\cite{musicbrainz} and 16GB~\cite{stackoverflow}, respectively.


Table~\ref{tab:segments_table} showcases several key points. First, for some applications, such as MySQL and Memcached, a single large segment covers most of the physical memory, and therefore DS would eliminate most TLB misses~\cite{basu:efficient}. Nevertheless, other applications, such as RocksDB and MonetDB (running TPC-H and TPC-DS), exhibit a large number of segments, and hence would expose the majority of the TLB misses. 

Second, for most applications, the total number of segments and the number of segments needed for 99\% coverage are much higher than what the RMM work assumes. On average, even for the small $8$GB dataset, the total number of segments is $4\times$ higher, and the number of segments for 99\% coverage is almost an order of magnitude higher than the requirements for the applications evaluated in \cite{gandhi:range}. The total number of segments places a burden on the number of range TLB entries. For instance, Memcached, which exhibits a very simple memory layout, requires a range TLB of $32$ entries to remove almost all the TLB misses, although there is a single segment that covers almost $100$\% of the memory. The reason is that accesses to other segments evict the largest segment's entry. Hence, Table~\ref{tab:segments_table}'s last column represents the best-case range TLB coverage for each workload. The range TLB is a fully associative structure, because segment sizes vary, making the standard indexing for set-associative structures hard. The area/energy requirements of this fully associative structure alone could dwarf the area/energy footprint of simple MPUs on the low-power logic die~\cite{ahn:scalable, drumond:mondrian, gao:practical, pugsley:ndc}. 

Third, although there could be hundreds of segments, the associativity requirements for Section~\ref{sec:vm}'s $8$GB dataset indicate that associativity can be reduced to a small number. The reason is that although segments are not fully contiguous, the OS tends to cluster the segments (as shown in Figure~\ref{fig:scope}), and therefore nearby segments do not conflict with each other. 

\sisetup{round-precision=3,round-mode=figures,scientific-notation=false}%

As seen in Table~\ref{tab:segments_table}, some applications, such as RocksDB and Cassandra, exhibit an increase in the number of segments as the dataset grows, increasing the pressure in both the range TLB and the rest of RMM structures. For DIPTA, we measure the sensitivity of page conflicts to associativity as dataset scales, employing Section~\ref{sec:vm}'s methodology and tuning the workloads to utilize $16$GB and $32$GB. Space limitations preclude a graphical representation of the results, which resemble Table~\ref{tab:rate_overhead}: 
Conflicts drop more sharply between direct-mapped and 2-way associativity, whereas 4-way associativity practically removes all page conflicts. In all cases---8GB, 16GB, and 32GB---4 ways make the page conflict overhead less than $1$\% of a memory access in the worst case. The reason associativity requirements do not increase is that the OS clusters segments around few places (e.g., heap and mmap areas). The increase in nearby segments does not increase the conflicts. In other words, the number of conflicts is more closely related to the number of clustered areas than to the number of segments.

Note that our experiments privilege DS and RMMs as we employ TCMalloc's memory allocator, which coalesces segments when possible. For instance, employing Glibc's memory allocator generates more than 800 segments for Memcached~\cite{park:efficient}, while we only require a few tens of them (as also corroborated by prior work~\cite{basu:efficient,karakostas:redundant}).

Last, RMM replaces the conventional demand paging policy for eager paging to improve contiguity in physical memory. Additionally, the OS has to manage virtual memory at a variable-sized granularity, which may create the external fragmentation problem that plagued the first segment-based VM designs~\cite{denning:virtual}. In contrast, DIPTA makes less disruptive changes to the OS operation as it employs conventional page sizes and the default demand-paging policy.

\section{Related work}
\label{sec:related}


\textbf{Processing in memory.} Recent advancements in die-stacking have enabled the integration of a sizable amount of low-power logic into conventional DRAM chips~\cite{hmc, diram}, solving the density and cost problem of planar chips that combined processing and DRAM~\cite{draper:architecture, kang:flexram, kozyrakis:scalable, oskin:active}. Leveraging this technology, several domain-specific architectures have emerged. NDC~\cite{pugsley:ndc}, Tesseract~\cite{ahn:scalable}, and NDP~\cite{gao:practical} consider a network of MPU chips. Tesseract relies on message passing without virtual memory. NDC and NDP assume a global address space with conventional address translation. Ahn et al.~\cite{ahn:pim-enabled} execute host instructions next to the memory, while translation is performed on the CPU.



\textbf{Unified address space.} Industry and academia have been recently pushing towards unified virtual memory between CPUs and GPUs. Examples include AMD's HSA~\cite{rogers:amd} and Nvidia's Unified Memory~\cite{harris:unified}, along with academic publications~\cite{pichai:architectural, power:supporting}. Both \cite{power:supporting} and \cite{pichai:architectural} propose a TLB architecture to sustain the high translation throughput required for GPU cores. A recent study on IOMMU translation for integrated GPUs has shown that a TLB miss takes an order of magnitude longer than on CPU cores~\cite{vesely:observation}. 


\textbf{Improving TLB performance.} Several studies exploited the contiguity generated by the buddy allocator and the memory compactor. CoLT \cite{pham:colt}, clustered \cite{pham:increasing}, and sub-blocked \cite{talluri:surpassing} TLBs group multiple PTEs into a single TLB entry. Direct segments~\cite{basu:efficient} allows for an efficient mapping between a single virtual segment mapped contiguously in physical memory. Karakostas et al.~\cite{karakostas:redundant} propose a fully associative range-based TLB and page table to transparently exploit the available contiguity in the virtual and physical address spaces. Transparent Huge Pages~\cite{transparenthugepages} and libHugeTLBFS~\cite{lighugetlbfs} increase the TLB reach by mapping large regions to a single TLB entry.

\textbf{Reducing associativity of virtual memory.} We are not the first to exploit reducing the associativity of VM. Several degrees of page coloring---fixing a few bits from the virtual-to-physical map---were proposed in the past. MIPS R6000 used page coloring coupled with a small TLB to index the cache under tight latency constraints~\cite{taylor:tlb}. Page coloring has also been used for virtually-indexed physically-tagged caches~\cite{chiueh:eliminating} as an alternative to large cache associativities~\cite{gustafson:ibm} or page sizes~\cite{jouppi:architectural}. Back in the 70s, Alan Jay Smith advocated the usage of set-associative mappings for main memory---much like a cache---to simplify the page placement and replacement~\cite{smith:comparative} .

\section{Conclusion}
\label{sec:conclusion}

 Providing the emerging memory+logic chips with VM enables a whole new class of applications to run near memory, facilitates the interaction between CPUs and MPUs, improves programmability, and enforces protection. To efficiently support near-memory address translation, we observe that the historically important flexibility to map a virtual page to any page frame is largely unnecessary in today's servers. While limiting the associativity incurs no penalty, it can break the translate-then-fetch serialization if combined with careful data placement in the MPU's memory, in which case translation and data fetch are independently performed. We propose DIPTA, a fully distributed inverted page table which stores the translation information next to the data, ensuring that the translation and data fetch always complete together, completely eliminating the translation overhead. 

\IEEEpeerreviewmaketitle

\section*{Acknowledgment}

The authors would like to thank Atri Bhattacharyya, Alexandros Daglis, Mario Drumond, Arash Pourhabibi, Georgios Psaropoulos, Ryan Rakvic, and Marko Vasic for their precious comments and feedback, and the anonymous reviewers for their insightful comments. This work has been
partially funded by the SNSF under
grant P2ELP2\_161823, CE-EuroLab-4-HPC, and the Nano-Tera YINS. 




%



\bstctlcite{bstctl:etal, bstctl:nodash, bstctl:simpurl}
\bibliographystyle{IEEEtranS}

\begin{thebibliography}{10}
\providecommand{\url}[1]{#1}
\csname url@samestyle\endcsname
\providecommand{\newblock}{\relax}
\providecommand{\bibinfo}[2]{#2}
\providecommand{\BIBentrySTDinterwordspacing}{\spaceskip=0pt\relax}
\providecommand{\BIBentryALTinterwordstretchfactor}{4}
\providecommand{\BIBentryALTinterwordspacing}{\spaceskip=\fontdimen2\font plus
\BIBentryALTinterwordstretchfactor\fontdimen3\font minus
  \fontdimen4\font\relax}
\providecommand{\BIBforeignlanguage}[2]{{%
\expandafter\ifx\csname l@#1\endcsname\relax
\typeout{** WARNING: IEEEtranS.bst: No hyphenation pattern has been}%
\typeout{** loaded for the language `#1'. Using the pattern for}%
\typeout{** the default language instead.}%
\else
\language=\csname l@#1\endcsname
\fi
#2}}
\providecommand{\BIBdecl}{\relax}
\BIBdecl

\bibitem{facebook:linkbench}
``{Facebook LinkBench Benchmark},''
  \url{https://github.com/facebook/linkbench}.

\bibitem{symantec:getting}
``{Getting The Hang Of IOPS v1.3},''
  \url{http://www.symantec.com/connect/articles/getting-hang-iops-v13}.

\bibitem{google:tcmalloc}
``{Google Performance Tools},'' \url{https://code.google.com/p/gperftools/}.

\bibitem{lighugetlbfs}
``{libHugeTLBFS},'' \url{http://linux.die.net/man/7/libhugetlbfs}.

\bibitem{hmc}
``{Micron Hybrid Memory Cube},''
  \url{http://www.micron.com/products/hybrid-memory-cube}.

\bibitem{musicbrainz}
``{MusicBrainz},'' \url{http://musicbrainz.org/}.

\bibitem{oracle:timesten}
``{Oracle TimesTen},''
  \url{http://www.oracle.com/technetwork/database/database-technologies/timesten/overview/index.html}.

\bibitem{facebook:rocksdb}
``{RocksDB In Memory Workload Performance Benchmarks},''
  \url{https://github.com/facebook/rocksdb/wiki/RocksDB-In-Memory-Workload-Performance-Benchmarks}.

\bibitem{sap:hana}
``{SAP HANA},'' \url{http://hana.sap.com/abouthana.html}.

\bibitem{stackoverflow}
``{Stack Overflow},'' \url{https://archive.org/details/stackexchange/}.

\bibitem{diram}
``{Tezzaron DiRAM},'' \url{http://www.tezzaron.com/products/diram4-3d-memory/}.

\bibitem{transparenthugepages}
``{Transparent huge pages in 2.6.38},'' \url{https://lwn.net/Articles/423584/}.

\bibitem{nvidia:uva}
``{NVIDIA's Next Generation CUDA Compute Architecture: Fermi.}''
  \url{http://www.nvidia.com/content/pdf/fermi_white_papers/nvidia_fermi_compute_architecture_whitepaper.pdf},
  2009.

\bibitem{jedec:high}
\BIBentryALTinterwordspacing
``{High Bandwidth Memory (HBM) DRAM},'' 2013. [Online]. Available:
  \url{http://www.jedec.org/standards-documents/results/jesd235}
\BIBentrySTDinterwordspacing

\bibitem{ahn:scalable}
J.~Ahn, S.~Hong, S.~Yoo, O.~Mutlu, and K.~Choi, ``A scalable
  processing-in-memory accelerator for parallel graph processing,'' in
  \emph{Proceedings of the 2015 International Symposium on Computer
  Architecture}, 2015.

\bibitem{ahn:pim-enabled}
J.~Ahn, S.~Yoo, O.~Mutlu, and K.~Choi, ``{PIM}-enabled instructions: a
  low-overhead, locality-aware processing-in-memory architecture,'' in
  \emph{Proceedings of the 2015 International Symposium on Computer
  Architecture}, 2015.

\bibitem{arm:cortex-a7}
ARM, ``{Cortex-A7 Processor},'' \url{http://www.arm.com/products/
  processors/cortex-a/cortex-a7.php}.

\bibitem{barr:translation}
T.~W. Barr, A.~L. Cox, and S.~Rixner, ``Translation caching: skip, don't walk
  (the page table),'' in \emph{Proceedings of the 2010 International Symposium
  on Computer Architecture}, 2010.

\bibitem{barr:spectlb}
T.~W. Barr, A.~L. Cox, and S.~Rixner, ``{SpecTLB}: A mechanism for speculative
  address translation,'' in \emph{Proceedings of the 2011 International
  Symposium on Computer Architecture}, 2011.

\bibitem{barroso:case}
L.~A. Barroso and U.~H{\"{o}}lzle, ``The case for energy-proportional
  computing,'' \emph{{IEEE} Computer}, vol.~40, no.~12, pp. 33--37, 2007.

\bibitem{basu:efficient}
A.~Basu, J.~Gandhi, J.~Chang, M.~D. Hill, and M.~M. Swift, ``Efficient virtual
  memory for big memory servers,'' in \emph{Proceedings of the 2013
  International Symposium on Computer Architecture}, 2013.

\bibitem{basu:reducing}
A.~Basu, M.~D. Hill, and M.~M. Swift, ``Reducing memory reference energy with
  opportunistic virtual caching,'' in \emph{Proceedings of the 2012
  International Symposium on Computer Architecture}, 2012.

\bibitem{bhattacharjee:large-reach}
A.~Bhattacharjee, ``Large-reach memory management unit caches,'' in
  \emph{Proceedings of the 2013 International Symposium on Microarchitecture},
  2013.

\bibitem{bhattacharjee:shared}
A.~Bhattacharjee, D.~Lustig, and M.~Martonosi, ``Shared last-level {TLB}s for
  chip multiprocessors,'' in \emph{Proceedings of the 2011 International
  Symposium on High-Performance Computer Architecture}, 2011.

\bibitem{bronson:tao}
N.~Bronson, Z.~Amsden, G.~Cabrera, P.~Chakka, P.~Dimov, H.~Ding, J.~Ferris,
  A.~Giardullo, S.~Kulkarni, H.~C. Li, M.~Marchukov, D.~Petrov, L.~Puzar, Y.~J.
  Song, and V.~Venkataramani, ``{TAO:} facebook's distributed data store for
  the social graph,'' in \emph{Proceedings of the 2013 Annual Technical
  Conference}, 2013.

\bibitem{calder:predictive}
B.~Calder, D.~Grunwald, and J.~S. Emer, ``Predictive sequential associative
  cache,'' in \emph{Proceedings of the 1996 International Symposium on
  High-Performance Computer Architecture}, 1996.

\bibitem{cheng:virtual}
R.~Cheng, ``Virtual address cache in unix,'' in \emph{Proceedings of the Summer
  1987 USENIX Technical Conf.}, 1987.

\bibitem{chiueh:eliminating}
T.~Chiueh and R.~H. Katz, ``Eliminating the address translation bottleneck for
  physical address cache,'' in \emph{Proceedings of the 1992 International
  Conference on Architectural Support for Programming Languages and Operating
  Systems}, 1992.

\bibitem{intel:tlbs}
I.~Corporation, ``Tlbs, paging-structure caches and their invalation,''
  \emph{Intel Technical Report}, 2008.

\bibitem{couleur:shared-access}
\BIBentryALTinterwordspacing
J.~F. Couleur and E.~L. Glaser, ``Shared-access data processing system,'' 1968,
  {US} Patent 3,412,382. Available:
  \url{http://www.google.com.ar/patents/US3412382}
\BIBentrySTDinterwordspacing

\bibitem{david:asynchronized}
T.~David, R.~Guerraoui, and V.~Trigonakis, ``Asynchronized concurrency: The
  secret to scaling concurrent search data structures,'' in \emph{Proceedings
  of the 2015 International Conference on Architectural Support for Programming
  Languages and Operating Systems}, 2015.

\bibitem{denning:virtual}
P.~J. Denning, ``Virtual memory,'' \emph{ACM Comput. Surv.}, vol.~2, no.~3, pp.
  153--189, 1970.

\bibitem{draper:architecture}
J.~Draper, J.~Chame, M.~Hall, C.~Steele, T.~Barrett, J.~LaCoss, J.~Granacki,
  J.~Shin, C.~Chen, C.~W. Kang, I.~Kim, and G.~Daglikoca, ``The architecture of
  the diva processing-in-memory chip,'' in \emph{Proceedings of the 2002
  International Conference on Supercomputing}, 2002.

\bibitem{drumond:mondrian}
M.~Drumond, A.~Daglis, N.~Mirzadeh, D.~Ustiugov, J.~Picorel, B.~Falsafi,
  B.~Grot, and D.~Pnevmatikatos, ``The mondrian data engine,'' in
  \emph{Proceedings of the 2017 Annual International Symposium on Computer
  Architecture}, 2017.

\bibitem{ferdman:clearing}
M.~Ferdman, A.~Adileh, Y.~O. Ko{\c{c}}berber, S.~Volos, M.~Alisafaee,
  D.~Jevdjic, C.~Kaynak, A.~D. Popescu, A.~Ailamaki, and B.~Falsafi, ``Clearing
  the clouds: a study of emerging scale-out workloads on modern hardware,'' in
  \emph{Proceedings of the 2012 International Conference on Architectural
  Support for Programming Languages and Operating Systems}, 2012.

\bibitem{gandhi:range}
J.~Gandhi, V.~Karakostas, F.~Ayar, A.~Cristal, M.~D. Hill, K.~S. McKinley,
  M.~Nemirovsky, M.~M. Swift, and O.~S. Unsal, ``Range translations for fast
  virtual memory,'' \emph{{IEEE} Micro}, vol.~36, no.~3, pp. 118--126, 2016.

\bibitem{gao:practical}
M.~Gao, G.~Ayers, and C.~Kozyrakis, ``Practical near-data processing for
  in-memory analytics frameworks,'' in \emph{Proceedings of the 2015
  International Conference on Parallel Architecture and Compilation
  Techniques}, 2015.

\bibitem{gustafson:ibm}
R.~N. Gustafson and F.~J. Sparacio, ``{IBM} 3081 processor unit: Design
  considerations and design process,'' \emph{{IBM} Journal of Research and
  Development}, vol.~26, no.~1, pp. 12--21, 1982.

\bibitem{gutierrez:integrated}
A.~Gutierrez, M.~Cieslak, B.~Giridhar, R.~G. Dreslinski, L.~Ceze, and T.~N.
  Mudge, ``Integrated 3d-stacked server designs for increasing physical density
  of key-value stores,'' in \emph{Proceedings of the 2014 International
  Conference on Architectural Support for Programming Languages and Operating
  Systems}, 2014.

\bibitem{mammarlund:4th}
P.~Hammarlund, ``{4th Generation
  Intel~\textregistered~Core~\texttrademark~Processor, codenamed Haswell},''
  \url{http://www.hotchips.org/wp-content/uploads/hc_archives/hc25/HC25.80-Processors2-epub/HC25.27.820-Haswell-Hammarlund-Intel.pdf},
  2013.

\bibitem{haque:few-to-many}
M.~E. Haque, Y.~H. Eom, Y.~He, S.~Elnikety, R.~Bianchini, and K.~S. McKinley,
  ``Few-to-many: Incremental parallelism for reducing tail latency in
  interactive services,'' in \emph{Proceedings of the 2015 International
  Conference on Architectural Support for Programming Languages and Operating
  Systems}, 2015.

\bibitem{harris:unified}
M.~Harris, ``Unified memory in cuda 6,''
  \url{http://on-demand.gputechconf.com/supercomputing/2013/presentation/SC3120-Unified-Memory-CUDA-6.0.pdf},
  2013.

\bibitem{ho:efficient}
C.~Ho, S.~J. Kim, and K.~Sankaralingam, ``Efficient execution of memory access
  phases using dataflow specialization,'' in \emph{Proceedings of the 2015
  International Symposium on Computer Architecture}, 2015.

\bibitem{hsieh:accelerating}
K.~Hsieh, S.~Khan, N.~Vijaykumar, K.~K. Chang, A.~Boroumand, S.~Ghose, and
  O.~Mutlu, ``Accelerating pointer chasing in 3d-stacked memory: Challenges,
  mechanisms, evaluation),'' in \emph{Proceedings of the 2016 International
  Conference on Computer Design}, 2016.

\bibitem{jacob:look}
B.~L. Jacob and T.~N. Mudge, ``A look at several memory management units,
  tlb-refill mechanisms, and page table organizations,'' in \emph{Proceedings
  of the 1998 International Conference on Architectural Support for Programming
  Languages and Operating Systems}, 1998.

\bibitem{jacob:virtual}
B.~L. Jacob and T.~N. Mudge, ``Virtual memory: Issues of implementation,''
  \emph{{IEEE} Computer}, vol.~31, no.~6, pp. 33--43, 1998.

\bibitem{jeon:predictive}
M.~Jeon, S.~Kim, S.~Hwang, Y.~He, S.~Elnikety, A.~L. Cox, and S.~Rixner,
  ``Predictive parallelization: taming tail latencies in web search,'' in
  \emph{Proceedings of the 2014 International {ACM} {SIGIR} Conference on
  Research and Development in Information Retrieval}, 2014.

\bibitem{jevdjic:unison}
D.~Jevdjic, G.~H. Loh, C.~Kaynak, and B.~Falsafi, ``Unison cache: {A} scalable
  and effective die-stacked {DRAM} cache,'' in \emph{Proceedings of the 2014
  Annual International Symposium on Microarchitecture}, 2014.

\bibitem{jouppi:architectural}
N.~P. Jouppi, ``Architectural and organizational tradeoffs in the design of the
  multititan cpu,'' in \emph{Proceedings of the 1989 International Symposium on
  Computer Architecture}, 1989.

\bibitem{jouppi:improving}
N.~P. Jouppi, ``Improving direct-mapped cache performance by the addition of a
  small fully-associative cache and prefetch buffers,'' in \emph{Proceedings of
  the 1990 International Symposium on Computer Architecture}, 1990.

\bibitem{kang:flexram}
Y.~Kang, W.~Huang, S.~Yoo, D.~Keen, Z.~Ge, V.~V. Lam, J.~Torrellas, and
  P.~Pattnaik, ``Flexram: Toward an advanced intelligent memory system,'' in
  \emph{Proceedings of the 1999 International Conference On Computer Design,
  {VLSI} in Computers and Processors}, 1999.

\bibitem{kanter:cavium}
\BIBentryALTinterwordspacing
D.~Kanter, ``{Cavium Thunders Into Servers: Specialized Silicon Rivals Xeon for
  Specific Workloads},'' 2016. Available:
  \url{http://www.linleygroup.com/mpr/article.php?url=mpr/h/2016/11545/11545.pdf}
\BIBentrySTDinterwordspacing

\bibitem{karakostas:redundant}
V.~Karakostas, J.~Gandhi, F.~Ayar, A.~Cristal, M.~D. Hill, K.~S. McKinley,
  M.~Nemirovsky, M.~M. Swift, and O.~\"{U}nsal, ``Redundant memory mappings for
  fast access to large memories,'' in \emph{Proceedings of the 2015
  International Symposium on Computer Architecture}, 2015.

\bibitem{karakostas:performance}
V.~Karakostas, O.~S. Unsal, M.~Nemirovsky, A.~Cristal, and M.~M. Swift,
  ``Performance analysis of the memory management unit under scale-out
  workloads,'' in \emph{Proceedings of the 2015 International Symposium on
  Workload Characterization}, 2014.

\bibitem{kim:memory-centric}
G.~Kim, J.~Kim, J.~H. Ahn, and J.~Kim, ``Memory-centric system interconnect
  design with hybrid memory cubes,'' in \emph{Proceedings of the 2013
  International Conference on Parallel Architectures and Compilation
  Techniques}, 2013.

\bibitem{kocberber:meet}
Y.~O. Ko{\c{c}}berber, B.~Grot, J.~Picorel, B.~Falsafi, K.~T. Lim, and
  P.~Ranganathan, ``Meet the walkers: accelerating index traversals for
  in-memory databases,'' in \emph{Proceedings of the 2013 International
  Symposium on Microarchitecture}, 2013.

\bibitem{kozyrakis:server}
C.~E. Kozyrakis, A.~Kansal, S.~Sankar, and K.~Vaid, ``Server engineering
  insights for large-scale online services,'' \emph{{IEEE} Micro}, vol.~30,
  no.~4, pp. 8--19, 2010.

\bibitem{kozyrakis:scalable}
C.~E. Kozyrakis, S.~Perissakis, D.~A. Patterson, T.~E. Anderson, K.~Asanovic,
  N.~Cardwell, R.~Fromm, J.~Golbus, B.~Gribstad, K.~Keeton, R.~Thomas,
  N.~Treuhaft, and K.~A. Yelick, ``Scalable processors in the
  billion-transistor era: {IRAM},'' \emph{{IEEE} Computer}, vol.~30, no.~9,
  1997.

\bibitem{lim:thin}
K.~T. Lim, D.~Meisner, A.~G. Saidi, P.~Ranganathan, and T.~F. Wenisch, ``Thin
  servers with smart pipes: designing soc accelerators for memcached,'' in
  \emph{Proceedings of the 2013 International Symposium on Computer
  Architecture}, 2013.

\bibitem{fujitsu:while}
F.~Limited, ``{White paper FUJITSU Supercomputer PRIMEHPC FX100 Evolution to
  the Next Generation},''
  \url{https://www.fujitsu.com/global/Images/primehpc-fx100-hard-en.pdf}, 2014.

\bibitem{luk:pin}
C.~Luk, R.~S. Cohn, R.~Muth, H.~Patil, A.~Klauser, P.~G. Lowney, S.~Wallace,
  V.~J. Reddi, and K.~M. Hazelwood, ``Pin: building customized program analysis
  tools with dynamic instrumentation,'' in \emph{Proceedings of the 2005
  Conference on Programming Language Design and Implementation}, 2005.

\bibitem{mars:bubble-up}
J.~Mars, L.~Tang, R.~Hundt, K.~Skadron, and M.~L. Soffa, ``Bubble-up:
  increasing utilization in modern warehouse scale computers via sensible
  co-locations,'' in \emph{Proceedings of the 2011 International Symposium on
  Microarchitecture}, 2011.

\bibitem{mauerer:professional}
W.~Mauerer, \emph{Professional Linux Kernel Architecture}.\hskip 1em plus 0.5em
  minus 0.4em\relax Birmingham, UK, UK: Wrox Press Ltd., 2008.

\bibitem{mcmillan:data}
R.~McMillan, ``{Data center servers suck - but nobody knows how much.}'' 2012.

\bibitem{micron:hmc}
\BIBentryALTinterwordspacing
Micron, ``{Hybrid Memory Cube Specification 2.1},'' 2014. Available:
  \url{http://www.hybridmemorycube.org/files/SiteDownloads/HMC-30G-VSR_HMCC_Specification_Rev2.1_20151105.pdf}
\BIBentrySTDinterwordspacing

\bibitem{oskin:active}
M.~Oskin, F.~T. Chong, and T.~Sherwood, ``Active pages: {A} computation model
  for intelligent memory,'' in \emph{Proceedings of the 1998 International
  Symposium on Computer Architecture}, 1998.

\bibitem{ousterhout:case}
J.~K. Ousterhout, P.~Agrawal, D.~Erickson, C.~Kozyrakis, J.~Leverich,
  D.~Mazi{\`{e}}res, S.~Mitra, A.~Narayanan, G.~M. Parulkar, M.~Rosenblum,
  S.~M. Rumble, E.~Stratmann, and R.~Stutsman, ``The case for ramclouds:
  scalable high-performance storage entirely in {DRAM},'' \emph{Operating
  Systems Review}, vol.~43, no.~4, pp. 92--105, 2009.

\bibitem{papadopoulou:prediction-based}
M.-M. Papadopoulou, X.~Tong, A.~Seznec, and A.~Moshovos, ``{Prediction-based
  superpage-friendly {TLB} designs},'' in \emph{Proceedings of the 2015
  symposium on High Performance Computer Architecture}, 2015.

\bibitem{parashar:triggered}
A.~Parashar, M.~Pellauer, M.~Adler, B.~Ahsan, N.~C. Crago, D.~Lustig,
  V.~Pavlov, A.~Zhai, M.~Gambhir, A.~Jaleel, R.~L. Allmon, R.~Rayess,
  S.~Maresh, and J.~S. Emer, ``Triggered instructions: a control paradigm for
  spatially-programmed architectures,'' in \emph{Proceedings of the 2013
  International Symposium on Computer Architecture}, 2013.

\bibitem{park:efficient}
H.~H. Park, H.~Taekyung, and J.~Huh, ``Efficient synonym filtering and scalable
  delayed translation for hybrid virtual caching,'' in \emph{Proceedings of the
  2016 International Symposium on Computer Architecture}, 2016.

\bibitem{pham:increasing}
B.~Pham, A.~Bhattacharjee, Y.~Eckert, and G.~H. Loh, ``Increasing {TLB} reach
  by exploiting clustering in page translations,'' in \emph{Proceedings of the
  2014 International Symposium on High Performance Computer Architecture},
  2014.

\bibitem{pham:colt}
B.~Pham, V.~Vaidyanathan, A.~Jaleel, and A.~Bhattacharjee, ``Colt: Coalesced
  large-reach {TLB}s,'' in \emph{Proceedings of the 2012 International
  Symposium on Microarchitecture}, 2012.

\bibitem{pichai:architectural}
B.~Pichai, L.~Hsu, and A.~Bhattacharjee, ``Architectural support for address
  translation on gpus: designing memory management units for cpu/gpus with
  unified address spaces,'' in \emph{Proceedings of the 2014 International
  Conference on Architectural Support for Programming Languages and Operating
  Systems}, 2014.

\bibitem{powell:reducing}
M.~D. Powell, A.~Agarwal, T.~N. Vijaykumar, B.~Falsafi, and K.~Roy, ``Reducing
  set-associative cache energy via way-prediction and selective
  direct-mapping,'' in \emph{Proceedings of the 2001 International Symposium on
  Microarchitecture}, 2001.

\bibitem{power:supporting}
J.~Power, M.~D. Hill, and D.~A. Wood, ``Supporting x86-64 address translation
  for 100s of {GPU} lanes,'' in \emph{Proceedings of the 2014 International
  Symposium on High Performance Computer Architecture}, 2014.

\bibitem{power:implications}
J.~Power, Y.~Li, M.~D. Hill, J.~M. Patel, and D.~A. Wood, ``Implications of
  emerging 3d {GPU} architecture on the scan primitive,'' \emph{{SIGMOD}
  Record}, vol.~44, no.~1, pp. 18--23, 2015.

\bibitem{pugsley:ndc}
S.~H. Pugsley, J.~Jestes, H.~Zhang, R.~Balasubramonian, V.~Srinivasan,
  A.~Buyuktosunoglu, A.~Davis, and F.~Li, ``{NDC:} analyzing the impact of
  3d-stacked memory+logic devices on mapreduce workloads,'' in
  \emph{Proceedings of the 2014 International Symposium on Performance Analysis
  of Systems and Software}, 2014.

\bibitem{qureshi:fundamental}
M.~K. Qureshi and G.~H. Loh, ``Fundamental latency trade-off in architecting
  {DRAM} caches: Outperforming impractical sram-tags with a simple and
  practical design,'' in \emph{Proceedings of the 2012 Annual International
  Symposium on Microarchitecture}, 2012.

\bibitem{reinders:knights}
J.~Reinders, ``{Knights Corner: Your Path to Knights Landing},''
  \url{https://software.intel.com/sites/default/files/managed/e9/b5/Knights-Corner-is-your-path-to-Knights-Landing.pdf},
  2014.

\bibitem{rogers:amd}
P.~Rogers, ``{AMD heterogeneous Uniform Memory Access.}''
  \url{http://events.csdn.net/AMD/130410%20-%20hUMA_v6.6_FINAL.PDF}.

\bibitem{saulsbury:recently-based}
A.~Saulsbury, F.~Dahlgren, and P.~Stenstr\"{o}m, ``Recency-based {TLB}
  preloading,'' in \emph{Proceedings of the 2000 Annual International Symposium
  on Computer Architecture}, 2000.

\bibitem{shevgoor:quantifying}
M.~Shevgoor, J.-S. Kim, N.~Chatterjee, R.~Balasubramonian, A.~Davis, and A.~N.
  Udipi, ``Quantifying the relationship between the power delivery network and
  architectural policies in a 3d-stacked memory device,'' in \emph{Proceedings
  of the 2013 International Symposium on Microarchitecture}, 2013.

\bibitem{smith:comparative}
A.~J. Smith, ``A comparative study of set associative memory mapping algorithms
  and their use for cache and main memory,'' \emph{{IEEE} Trans. Software
  Eng.}, vol.~4, no.~2, pp. 121--130, 1978.

\bibitem{talluri:surpassing}
M.~Talluri and M.~D. Hill, ``Surpassing the {TLB} performance of superpages
  with less operating system support,'' in \emph{Proceedings of the 1994
  International Conference on Architectural Support for Programming Languages
  and Operating Systems}, 1994.

\bibitem{taylor:tlb}
G.~Taylor, P.~Davies, and M.~Farmwald, ``The tlb slice---a low-cost high-speed
  address translation mechanism,'' in \emph{Proceedings of the 1990
  International Symposium on Computer Architecture}, 1990.

\bibitem{towles:unifying}
B.~Towles, J.~P. Grossman, B.~Greskamp, and D.~E. Shaw, ``Unifying on-chip and
  inter-node switching within the anton 2 network,'' in \emph{Proceedings of
  the 2014 International Symposium on Computer Architecture}, 2014.

\bibitem{vesely:observation}
J.~Vesel{\'{y}}, A.~Basu, M.~Oskin, G.~H. Loh, and A.~Bhattacharjee,
  ``Observations and opportunities in architecting shared virtual memory for
  heterogeneous systems,'' in \emph{Proceedings of the 2016 International
  Symposium on Performance Analysis of Systems and Software}, 2016.

\bibitem{villavieja:didi}
C.~Villavieja, V.~Karakostas, L.~Vilanova, Y.~Etsion, A.~Ram{\'{\i}}rez,
  A.~Mendelson, N.~Navarro, A.~Cristal, and O.~S. Unsal, ``Didi: Mitigating the
  performance impact of {TLB} shootdowns using a shared {TLB} directory,'' in
  \emph{Proceedings of the 2011 International Conference on Parallel
  Architectures and Compilation Techniques}, 2011.

\bibitem{volos:fat}
S.~Volos, D.~Jevdjic, B.~Falsafi, and B.~Grot, ``Fat caches for scale-out
  servers,'' \emph{IEEE Micro}, 2016.

\bibitem{wenisch:simflex}
T.~F. Wenisch, R.~E. Wunderlich, M.~Ferdman, A.~Ailamaki, B.~Falsafi, and J.~C.
  Hoe, ``Simflex: Statistical sampling of computer system simulation,''
  \emph{{IEEE} Micro}, vol.~26, no.~4, pp. 18--31, 2006.

\bibitem{wu:navigating}
L.~Wu, R.~J. Barker, M.~A. Kim, and K.~A. Ross, ``Navigating big data with
  high-throughput, energy-efficient data partitioning,'' in \emph{Proceedings
  of the 2013 International Symposium on Computer Architecture}, 2013.

\bibitem{xi:beyond}
S.~L. Xi, O.~Babarinsa, M.~Athanassoulis, and S.~Idreos, ``Beyond the wall:
  Near-data processing for databases,'' in \emph{Proceedings of the 2015
  International Workshop on Data Management on New Hardware (DaMoN)}, 2015.

\bibitem{yang:bubble-flux}
H.~Yang, A.~D. Breslow, J.~Mars, and L.~Tang, ``Bubble-flux: precise online qos
  management for increased utilization in warehouse scale computers,'' in
  \emph{Proceedings of the 2013 International Symposium on Computer
  Architecture}, 2013.

\bibitem{yoon:revisiting}
H.~Yoon and G.~S. Sohi, ``Revisiting virtual {L1} caches: {A} practical design
  using dynamic synonym remapping,'' in \emph{Proceedings of the 2016
  International Symposium on High Performance Computer Architecture}, 2016.

\bibitem{zaharia:spark}
M.~Zaharia, M.~Chowdhury, M.~J. Franklin, S.~Shenker, and I.~Stoica, ``Spark:
  Cluster computing with working sets,'' in \emph{Proceedings of the 2010
  Conference on Hot Topics in Cloud Computing}, 2010.

\end{thebibliography}

\end{document}